\documentclass[11pt]{article}
\PassOptionsToPackage{table}{xcolor}
\PassOptionsToPackage{hyphens}{url}

\usepackage[letterpaper,margin=1in]{geometry}
\usepackage{comment}
\usepackage{natbib}
\usepackage[font=small,labelfont=bf,skip=6pt]{caption}
\usepackage{authblk}

\usepackage{hyperref}
\hypersetup{colorlinks=true, breaklinks=true,
            linkcolor=[rgb]{0.00,0.20,0.55},   
            citecolor=[rgb]{0.00,0.35,0.15},   
            urlcolor=[rgb]{0.00,0.20,0.70}}
\usepackage{latexsym}
\usepackage[T1]{fontenc}
\usepackage[utf8]{inputenc}
\usepackage{microtype}
\usepackage{newunicodechar}
\newunicodechar{’}{'}
\newunicodechar{‘}{`}
\newunicodechar{“}{``}
\newunicodechar{”}{''}
\newunicodechar{—}{---}
\newunicodechar{–}{--}

\usepackage{graphicx}
\usepackage{amsmath}
\usepackage{booktabs}
\usepackage[table]{xcolor}
\usepackage{comment}
\usepackage{float}
\usepackage{placeins}
\usepackage{subcaption}
\usepackage{listings}
\usepackage{multirow}

\usepackage{algorithm}        
\usepackage{algpseudocode}    

  \usepackage[most]{tcolorbox}
  \tcbuselibrary{breakable}

  \usepackage{amssymb}    

  \usepackage{colortbl}
  \definecolor{cellyes}{HTML}{C8E6C9}                                                                          
  \definecolor{cellno}{HTML}{FFCDD2}    
  \definecolor{cellnull}{HTML}{E0E0E0}                                    
                                                                                                                             
  \newcommand{\y}{\cellcolor{cellyes}\checkmark}                                                                              
  \newcommand{\n}{\cellcolor{cellno}$\times$}                                                                                 
  \newcommand{\nul}{\cellcolor{cellnull}$\cdot$}                                                                                                                                          
  \newtcolorbox{promptbox}[1]{                                                                                                                     
    breakable,                                                                                                                                     
    colback=black!3,                                                                                                                               
    colframe=black!30,                                                                                                                             
    boxrule=0.4pt,                                                                                                                                 
    left=10pt, right=10pt, top=8pt, bottom=8pt,
    title=#1,                                                                                                                                      
    fonttitle=\small\bfseries,            
    coltitle=black,                                                                                                                                
    colbacktitle=black!8,                                                                                                                          
    enhanced,                                 
    attach boxed title to top left={xshift=8pt, yshift=-3pt},                                                                                      
    boxed title style={colback=black!8, colframe=black!30, boxrule=0.4pt},
    fontupper=\small,                                                                                                                              
    parbox=false,                             
    before upper={\setlength{\parindent}{0pt}\setlength{\parskip}{4pt plus 1pt}},                                                                  
  }                     
\usepackage{tikz}
\graphicspath{{figures/}}

\definecolor{methodpurple}{RGB}{128,0,128}
\definecolor{methodblue}{RGB}{0,102,153}

\title{Representing license permissiveness with partial orders}
\author[1]{Hamidah Oderinwale*}
\author[2]{David Atkinson}
\author[3]{Rachel Hong}
\author[4]{Art Abal}
\author[5]{Ben Laufer}
\affil[1]{McGill University \quad \texttt{hamidah.oderinwale@mail.mcgill.ca}}
\affil[2]{Georgetown University \quad \texttt{daa80@georgetown.edu}}
\affil[3]{University of Washington \quad \texttt{hongrach@cs.washington.edu}}
\affil[4]{Open Data Labs \quad \texttt{art@opendatalabs.xyz}}
\affil[5]{Cornell Tech \quad \texttt{bdl56@cornell.edu}}
\begin{document}
\maketitle
\begin{abstract}
Licenses are legal instruments that inventors rely upon to protect the technologies they build and regulate how they are used---however, the nature of their authorship and selection implies that how they are interpreted, chosen, and enforced is largely unstructured. In practice, this makes it difficult to compare licenses at scale---when is one license considered more permissive than the other, and when are their terms incomparable to each other?  Currently, there is a growing list of licenses that are introduced and used, yet no systematic way to study their relationships. This matters for platforms such as Hugging Face, GitHub, and the Python Package Index, where developers publish or build upon technologies that each have their own licenses. Using large language models (LLMs), we introduce methods for comparing  licenses at scale: first, in a pairwise fashion to construct and validate a partial ordering based on permissiveness; and by drawing on existing taxonomies of software licenses. Then, we try to recover the structure with the Bradley--Terry model to see if permissiveness can be judged more cheaply and observe a loss of ~20\%--and classify this loss to feature coverage. The former coupled with model rationale allows us to trace restrictiveness, and the latter allows us to understand license selection as a combination of shared provisions.  Our repository and data are available at:
\url{https://github.com/hamidahoderinwale/software-license-ordering}.

\end{abstract}

\section{Introduction}



Incentives for innovation often hinge on a developer's ability to protect the technologies they build \citep{scotchmer2004innovation}. Yet today's legal instruments have not been reconsidered in light of an increasingly AI-driven software ecosystem---where new kinds of software can be deployed with much more ease and at a faster pace \citep{Hopkins_2025_supply_chain}. The software ecosystem is built on dependencies, where libraries are imported into repositories and foundation models coupled with specialized datasets are deployed to accomplish specialized tasks \citep{laufer2025anatomymachinelearningecosystem}. When the developer of a technology chooses a license, they adopt its terms. When a user uses the software for which the license is chosen, they enter into an agreement to abide by those terms \citep{SaintLaurent2004UnderstandingOS}. If terms require attribution or that derivatives be released under the same license as the source, ignoring that requirement may constitute a breach of contract \citep{duan2025position}. The authorship of licenses is an interesting case of the commons-based peer production. We present methods for studying the structure of the software license ecosystem as a network.

\subsection{Toward (computable) software license governance}

While you can consider a software license a contract because it \emph{specifies terms} under which users may use a given software, there is no established structure on what terms can be written or how. As a corollary, the mechanisms by which these terms are enforced are similarly ill-defined. Licenses are unique because how they are written and published is largely unstructured. 
We extract clauses from licenses based on their headers to report their depth and breadth as has been done for U.S. code which are hierarchical by design \cite{Jeong2026}. 50.8\%, of the licenses have no nesting at all and the maximum depth observed in the other half is three. Licenses are also multi-purpose, written for technologies that may not presently exist. 

We focus primarily on the relicensing problem---where licenses are considered in terms of how they interact within composite software. We not only consider how they interact in composite technologies but also how they behave when there are dependencies. More formally, given an upstream license, we consider the set of compatible licenses available. 


New licenses are written for several reasons, but enforcement mechanisms are rarely described within them and typically have to be inferred. For example, while there are often terms of use which govern how an end user interacts or accesses the technology, it is less clear what terms apply to a derivative developer building on that technology who wishes to apply their own protections. Take, for instance, OpenClaw, an open-source agent project released with intentions to be a user-controlled, sovereign technology. OpenClaw comes with an MIT license which meant companies like OpenAI could adapt and release the project under more closed terms that may diverge from the creators' original intentions \citep{csharpcornerOpenClawReally}.

Given the arbitrariness of practices within the landscape, downstream artifacts do not always comply with the licenses of the artifacts they are built on. Software licenses are considered contracts and should be treated as such \citep{jacobsen_v_katzer}, which can lead to potential legal consequences. In 2022, the class-action lawsuit DOE 1 v. GitHub alleged that the creation of the Codex and Copilot AI systems constituted a breach of contract of open-source software licenses \citep{knowingmachinesGitHub}. This legal challenge implies that downstream licenses may also need to be compatible with upstream license restrictions, yet the current free-for-all of license attachments challenges our ability to make such comparisons.

\subsection{Contributions}

Based on these motivations, this work proposes methodology for reasoning about licenses and unstructured legal texts more broadly, and demonstrates its use in practice. To do so, we:
\begin{itemize}
    \item Present a dataset of 2,226 unique licenses: 747 are retrieved from SPDX (the Software Package Data Exchange), of which 93 are selectable on the Hugging Face license menu; 1,479 are retrieved from Hugging Face model cards. Each license in the dataset includes their text and metadata (author, version, etc.) and use existing software license taxonomies such as Nordlander's \citep{nordlander2004} and Kapitsaki's \citep{kapitsaki_open_source} to characterize them. 
    \item Employ pairwise comparisons to produce a partial ordering of licenses based on their permissiveness, and demonstrate that the Bradley-Terry algorithm is not a reliable way to recover the total-ordering when new licenses are added by their feature-predicted permissiveness ``scores.'' We observe that features are relatively unreliable predictors as a weighted sum.
  \item Analyze the OSS ecosystem across software library ecosystems covering a total of 9.3M packages from 2010--2025 (npm, PyPI, Cargo, Maven, and conda-forge), and find that inconsistent license choice---where a downstream artifact uses a more permissive license than an upstream dependency---happens at rates ranging from 12.3\% to 57.4\%.
\end{itemize}

\section{Related Work}

Licenses are the primary instrument through which innovators govern access to the technologies they develop, though software and new kinds of digital artifact exist in a gray zone of governance \citep{reuel2025openproblemstechnicalai}. This raises the question of whether license choice within a technological ecosystem can be optimally designed, and how the externalities of the choices made by its actors can be measured \citep{jewitt2026permissivewashingopenaisupply, HellerEisenberg1998Anticommons}. \citet{white2024modelopennessframeworkpromoting} put forth a framework for assessing openness and the relationships of artifacts in the model ecosystem. Others have tried to map the interactions of technology developers, in particular the AI supply chain---where interactions are particularly sprawling due to the diversity of interacting components \citep{Hopkins_2025_supply_chain}---and have empirically documented the license landscape across the ecosystem of ML models \citep{longpre2023dataprovenanceinitiativelarge, laufer2025anatomymachinelearningecosystem}.

\section{A dataset of software licenses}

Together, we collected 996 text-bearing licenses, filtering from 2,226 total  licenses with unique IDs: 747 are sourced from SPDX and 1,479 additional identifiers recovered from Hugging Face model cards and metadata of which 249 are actually text-bearing. We check for identical licenses by checking if they are `byte-identical' and find that 83 are redundant and 1,230 do not have any text (55.3\%).\footnote{A list of these licenses can be found in our dataset repository. A license version variant is a license with a version number in its SPDX ID.} For every license, we retrieve its text, its source, and any additional metadata offered by the provider. The majority of the licenses had their data retrieved from the Software Package Data Exchange (SPDX) \citep{spdx-3-0-1} started in 2011 which is a repository of indexed licenses in XML. We find that a small number of license are adopted by the majority, of the  637,980 models with licenses on Hugging Face, 11 licenses make up 90\% of the ones selected \ref{fig:license_model_share_hf}.\footnote{As calculated from a Hugging Face snapshot from July 31, 2026 of 2.95M models.}


\subsection{Who writes licenses?}
A sizeable fraction of licenses are written by non-profits and foundations such as the Creative Commons and the Free Software Foundation (FSF); both have relatively large (55\footnote{We combine counts from Creative Commons (49) and the Creative Commons Corporation (6).} and 66, respectively) and popular suites of licenses. The second largest demographic of license authors are individuals, examples including Donald Arseneau, the author of a number of LaTeX packages and three related licenses (e.g. the Dotseqn introduced in 1995); Henry Spencer, developer of regex libraries and the author of one corresponding license (the Spencer license), and Larry Wall, the developer of the Perl programming language and the author of the Artistic License family \citep{brand1994how, houston_regex, artistic_license}. 

The author dynamics of individuals reinforce Choksi and Grimmelmann's thesis, built on ideas from Brand, which presents licenses and infrastructure more broadly are built in live ecosystems and out of adapting needs \citep{ChoksiGrimmelmann2024HowLicensesLearn, brand1994how}. To classify organization types we use Wikidata's taxonomy. For the breakdown of author demographics we use the Research Organization Registry (ROR) which consists of nine categories and we add a category to categorize individuals as authors. Out of the ten, eight are represented, as archive and funder organizations were not classified as authors in the dataset, but we identify them as relatively ambiguous types.
  

\section{Arriving at a partial ordering}

We introduce methods for assessing the relative permissiveness of licenses by using LLMs to produce a partial ordering over the licenses in our set \citep{gu2025surveyllmasajudge}.   For every pair from a total of $4{,}278$ ($\binom{93}{2}$), we instruct
  models to select the more permissive license out of the two, or identify
  them as equal or incomparable. We use a subset of the licenses selectable by developers on Hugging Face because of the computational costs of calculating every pair ($\binom{996}{2}$) (495,510) from a broader set.

\subsubsection*{Formal definitions}
\label{sec:formal-definitions}

To denote the relationships more formally, let $L = \{\ell_1, \ell_2, \dots, \ell_n\}$ denote the set of licenses in the ML ecosystem;
our goal is to characterize the relationships between these licenses. We define two relations:
the first, represents a relationship where one license $\ell_i$ is therefore more permissive (less strict) than another license $\ell_j$: \begin{equation}
\ell_i > \ell_j
\end{equation} where $\ell_i$ $\rightarrow$ $\ell_j$ in a graph. Strictness is defined by a logical entailment relationship: If $\ell_i > \ell_j$ then any work that complies with $\ell_j$ must also comply with $\ell_i$, and there exists works that comply with $\ell_i$ but not $\ell_j$. The second represents a relationship where both licenses are equally as restrictive:   \begin{equation}
  \ell_i \sim \ell_j                        
  \end{equation} The third, is the lack of relation where both a pair is deemed incomparable due to a lack of consensus or certainty. 

\begin{equation}
  \ell_i \parallel \ell_j  
  \end{equation}

This is the most ambiguous relation, and we query models to select this choice in addition to one of the above so we can analyze its determinants and the nature of conflicting verdicts across raters.

\begin{figure}
    \centering
    \includegraphics[width=0.3\linewidth]{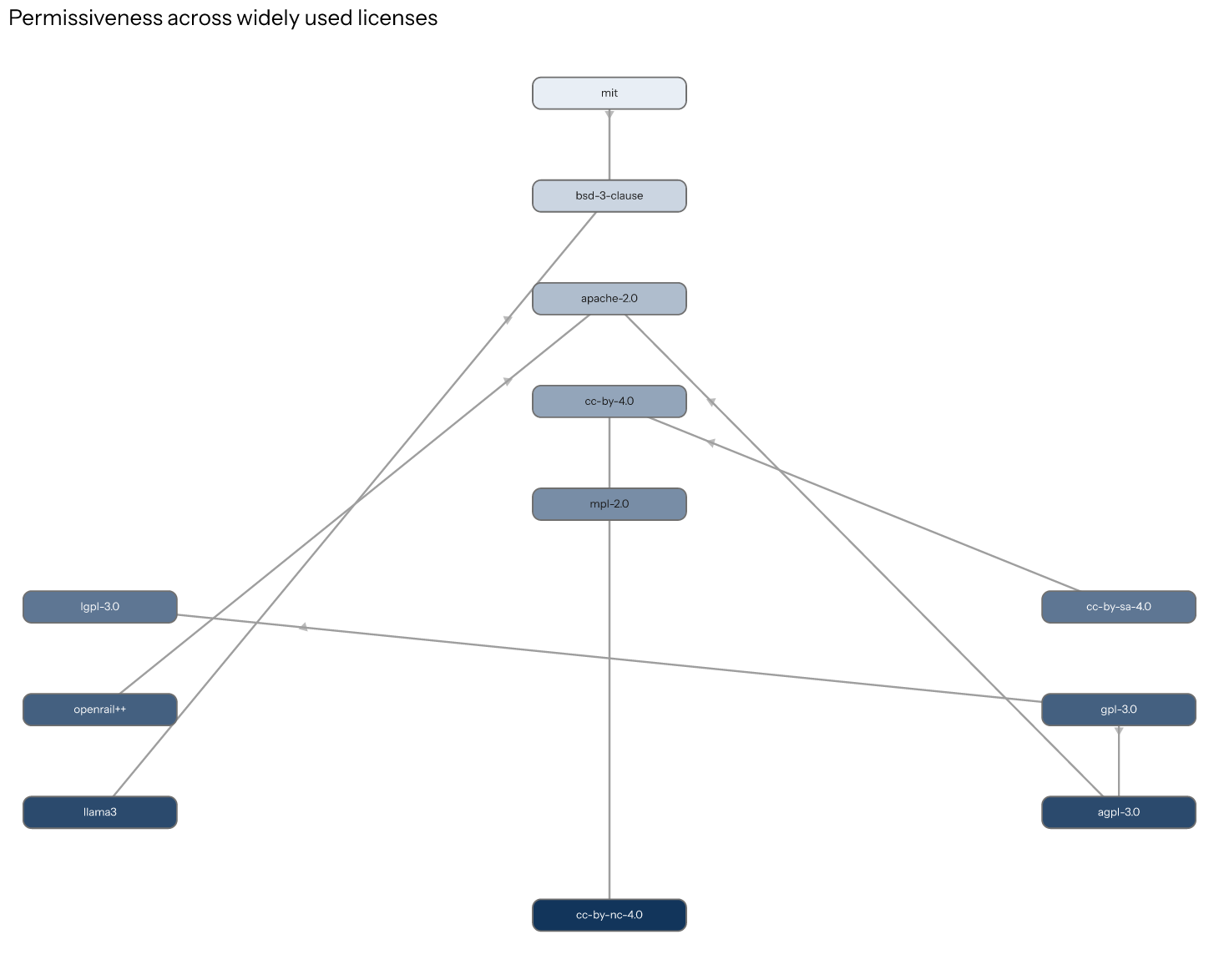}
    \caption{Partial-order diagram of a subset of 12 partially ordered licenses (incomparable pairs are left unconnected), showing the transition from the MIT license (most permissive) to the cc-by-nc-4.0 license (most restrictive). }
    \label{fig:partial-order-diagram}
\end{figure}

\section{Comparatively characterizing licenses}
\subsection{How license text changes with permissiveness}

Given the partial orderings we produce, we explore the relationship between textual features and permissiveness. Past work has investigated the complexity of legal literature over time \citep{Jeong2026-op} and has found that legal texts have grown more complex and long, with vocabularies staying increasingly dominated by a few popular terms. We observe similar findings in our dataset looking at lexical features of licenses as they get more permissive. For every pair of licenses in which one is deemed more permissive than the other, we note whether a feature is true (`appears') or false (`disappears') as a license becomes permissive (is chosen as the more permissive one in the ordered pair).

We look at license word counts as they become more restrictive. For diversity---the linguistic novelty and variation of a license's text---we employ Shannon entropy \citep{shannonmathematicaltheoryofcommunication}, which studies how evenly distributed vocabulary terms are within a corpus. We also compare embedding models for this task. Traditional models like MiniLM-L6-v2 (truncating ~70\% of the licenses) struggle with long texts \cite{mamakas2022processinglonglegaldocuments} so we employ two others for the same tasks (gte-modernbert-base and Qwen3-Embedding-0.6B).

Comparing these three we see that the models' representations of legal text do not represent permissiveness (as directed pairs are not meaningfully more similar than random pairs), however they do represent licenses families \ref{fig:embedding-atlas}. Possibly, this is because a license in the same family are likely to have shared textual quirks, while permissiveness is determined by specific clauses that can be difficult to discriminate because of how they are expressed. When comparing LLM pairwise orderings, we also see that there is a strong positive correlation between license popularity and agreement, agreement is raised from 91-94\% to 98.6\%-99.4\% when popularity is weighed. Potentially, this could be due to exposure during pretraining or user LLM interactions, or because popularity is a function of accessibility, where the most widely adopted licenses are also the clearest.


\begin{table*}[t]
    \centering\footnotesize
    \setlength{\tabcolsep}{2.6pt}
    \begin{tabular}{lrrrrrrrrr}
      \toprule
       & $n$ & \multicolumn{3}{c}{Cosine similarity} & TF-IDF & $\Delta$ words & $\Delta$ entropy & \multicolumn{2}{c}{$|\Delta|$ features} \\
      \cmidrule(lr){3-5}\cmidrule(lr){9-10}
       &  & MiniLM & ModernBERT & Qwen3 &  &  &  & Nord. & with Kap. \\
      \midrule
      All directed pairs           & 2{,}459   & 0.480 & 0.711 & 0.528 & 0.450 & $+689$ & $+0.48$ & 3.18 & 4.20 \\
      \rowcolor{black!6}
      All incomparable pairs       & 838       & 0.464 & 0.715 & 0.520 & 0.467 & $+1{,}729$ & $+0.45$ & 3.01 & 4.21 \\
      \midrule
      Within-family pairs          & 3{,}044   & 0.706 & 0.818 & 0.741 & 0.550 & $+477$ & $+0.43$ & 2.13 & 2.49 \\
      \rowcolor{black!6}
      Cross-family pairs           & 351{,}859 & 0.399 & 0.658 & 0.490 & 0.273 & $+1{,}147$ & $+0.99$ & 2.98 & 3.86 \\
      Random baseline              & 354{,}903 & 0.401 & 0.659 & 0.492 & 0.275 & $+1{,}142$ & $+0.99$ & 2.98 & 3.85 \\
      \bottomrule
    \end{tabular}
    \caption{Semantic and feature similarity across license pairs. Random and
    family-based pairs are sampled from the full corpus, while directed pairs
    are sampled from selectable licenses on Hugging Face. Semantic similarity is
    computed using cosine similarity from three embedding models:
    MiniLM-L6-v2, gte-modernbert-base, and Qwen3-Embedding-0.6B. For directed
    pairs, feature deltas are computed by subtracting the values of the more
    permissive license from the corresponding values of the less permissive
    license. Cross-family pairs contain licenses from different license
    families, while within-family pairs contain licenses from the same family
    (e.g., the Creative Commons family). The random baseline samples pairs
    without conditioning on ordering or family membership. Feature differences
    count the number of binary feature dimensions that differ between two
    licenses under the Nordlander taxonomy and the extended taxonomy including
    Kapitsaki features.}
    \label{tab:pairwise-comparisons}
\end{table*}



\subsection{License traits}
Notably, a shortcoming of the aforementioned is that it does not consider the nature of the license provisions themselves. For example, two licenses could have an equal number of claims but correspond to distinct areas of law. That is, a license could be more structurally restrictive but contextually incomparable. To address this, we extract binary features derived from existing taxonomies that characterize software licenses by their terms \citep{nordlander2004, kapitsaki_open_source}.
                 
\begin{table*}[t]
  \centering\footnotesize
  \setlength{\tabcolsep}{4pt}
  \begin{tabular}{@{}c|l>{\raggedright\arraybackslash}p{8.6cm}r@{}}
  \toprule
   & \textbf{Feature} & \textbf{Description} & \textbf{Licenses} \\
  \midrule
    & Attribution (Attr)      & Redistribution must preserve credit to the original author & 765 (91\%) \\
    \rowcolor{black!6}
    & ShareAlike (SA)         & Distributions must use the same or a compatible license & 233 (28\%) \\
    & Modification (Mod)      & Modification of the code is permitted & 807 (96\%) \\
    \rowcolor{black!6}
    & Derivatives (Der)       & Derivative works are permitted & 805 (95\%) \\
    & Commercial deriv.\ (CDer)  & Derivatives may be distributed commercially & 686 (81\%) \\
    \rowcolor{black!6}
    & Commercial redist.\ (CRed) & Redistribution may be charged for & 689 (82\%) \\
    & GPL-compatible (GPL)    & Can be combined with GPL code and distributed under the GPL & 245 (29\%) / 50 (43\%) \\
    \rowcolor{black!6}
    & Binaries (Bin)\textsuperscript{*} & Compiled binary distributions are permitted & 797 (95\%) \\
    & Source disclosure (Src) & Distributing modifications requires making source available & 155 (18\%) \\
    \rowcolor{black!6}
    \multirow{-10}{*}{\rotatebox{90}{Nordlander}}
    & Copyleft (CL)           & Any copyleft or ShareAlike requirement exists & 245 (29\%) \\
  \midrule
    & Patent use              & Explicit grant of patent rights to users & 153 (18\%) \\
    \rowcolor{black!6}
    & State changes           & Modifications must be documented when distributing & 320 (38\%) \\
    \multirow{-3}{*}{\rotatebox{90}{Kap.}}
    & Network use             & Network access triggers source-disclosure obligations & 38 (5\%) \\
  \bottomrule
  \end{tabular}
  \caption{License features from existing taxonomies
  \citep{nordlander2004,kapitsaki_open_source}, with the number of licenses exhibiting each feature (Opus~4.6). Features whose
  value cannot be determined from the license text alone are not counted which affects GPL-compatibility (220) and attribution (14) the most---extractable from identifiers.
  \textsuperscript{*}Binaries are compiled, machine-executable files
  (e.g.\ \texttt{.so}, \texttt{.exe}), distinct from human-readable source.}
  \label{tab:license_features}
\end{table*}

\subsection{How features correlate with permissiveness}

We look at how features change with permissiveness. Below, we plot the direction that each feature changes in as licenses become more restrictive. Out of the 1,213 ordered pairs where copyleft has a different value between the more permissive and less permissive license, it appears in the less permissive license 75\% of the time, showing a mean drift of +0.38. Given the absolute mean drift we can produce an ordering of features in the direction of restrictiveness. As seen in the plot, requiring attribution offers the strongest signal for restrictiveness while allowing commercial use is the strongest signal of permissiveness. 

\begin{figure}[t]      
\centering
\includegraphics[width=0.62\linewidth]{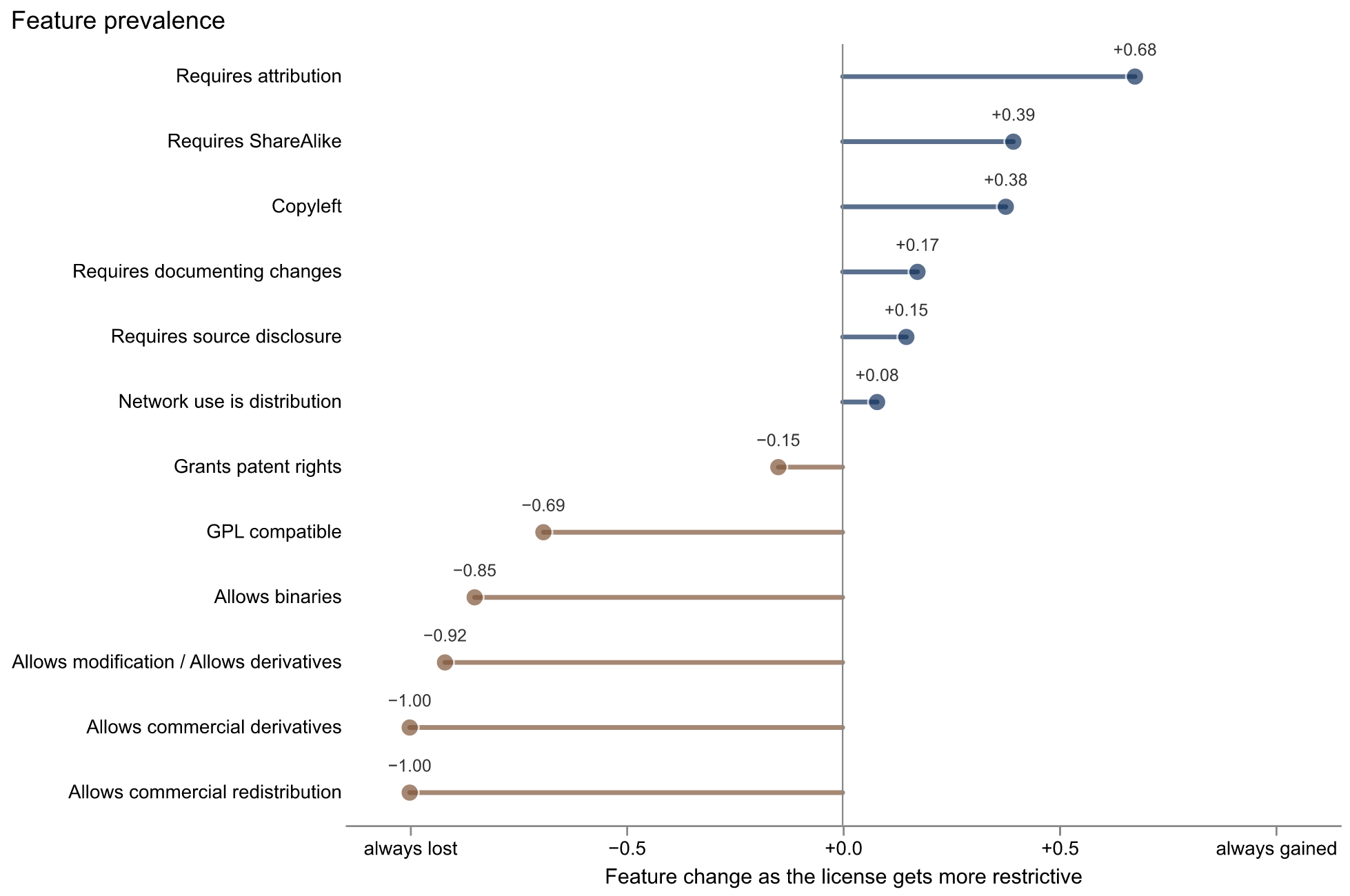} 
          \caption{Barbell plot showing how license features change in directed pairs, where licenses gets more restrictive. A feature gained is
  represented as +1 and a feature removed is represented as -1. We only count \emph{flips}, where a feature is either gained or lost in a directed
  pair. The plot shows the features that are associated with restriction in purple (positive values) and the features associated with              
  permissiveness (negative value) in blue.}\label{fig:feature_barbell}
  \end{figure} 


For a given feature, its order is determined by its score that is calculated accordingly:
   
  $$\text{mean direction} = \frac{\text{appears} - \text{disappears}}{\text{appears} + \text{disappears}}$$

With the feature profiles that we produce, we also categorize the most common feature changes as licenses get more restrictive. We group directed pairs by the number of features that differ. Then, we see the most common changes for each of them. For example, when there is a single feature change, the most common is GPL compatibility (50\%) followed by attribution (25\%).

With our features, we introduce a framework for comparing licenses by their functional signatures, that is, what they actually enforce. Each license is represented as a binary string, where every position corresponds to a proposition from our predefined vocabulary of features. Each position is evaluated depending on whether the assigned proposition is true or false.\footnote{For this framework, a feature needs to be a proposition: a claim that can be argued for or against. For example, ``commercial use'' is a topic, but ``derivatives of this license are allowed to be distributed for commercial use'' is a proposition because it is possible to verify whether the claim is true or false. More formally, we can describe a feature as a predicate that returns true or false \citep{MayoWilsonPredicateLogic}, where a lack of a positive determination, or an explicit description, is considered false. Nordlander's features follow the original taxonomy, while Kapitsaki's features were adapted as propositions.} While the partial ordering helps us understand the degree of restrictiveness across licenses, this framework
  allows us to compare licenses at a functional level (Table~\ref{tab:top_profiles}).

While single features can be directionally more permissive or restrictive, licenses are composed of multiple features and we also study them in combination. We can understand the ecosystem of licenses and what restrictions are most common by producing license profiles. With the 13 features from the Nordlander and Kapitsaki taxonomies, we characterize each license by its value for each feature in a string. Given these licenses, there is an upper bound of $2^{13}$ combinations (when null is not treated as a unique value, otherwise it is $3^{13}$). Yet, we find only 99 unique feature combinations when the features are considered with binary values (or 128 when they are considered with ternary values), with 57.5\% of licenses sharing the same 10 combinations.

\begin{table*}[t]
\centering
\footnotesize
\setlength{\tabcolsep}{4pt}
\renewcommand{\arraystretch}{1.2}

\resizebox{\textwidth}{!}{%
\begin{tabular}{@{}r *{13}{c} >{\raggedright\arraybackslash}p{5.0cm}@{}}
\toprule
& \multicolumn{10}{c}{\textbf{Nordlander}}
& \multicolumn{3}{c}{\textbf{Kapitsaki}}
& \\
\cmidrule(lr){2-11}\cmidrule(lr){12-14}

\textbf{Licenses}
& \textbf{Attr}
& \textbf{SA}
& \textbf{Mod}
& \textbf{Der}
& \textbf{CDer}
& \textbf{CRed}
& \textbf{GPL}
& \textbf{Bin}
& \textbf{Src}
& \textbf{CL}
& \textbf{Pat}
& \textbf{StCh}
& \textbf{Net}
& \textbf{Example licenses}
\\

\midrule

122 (14.5\%) 
& \y & \n & \y & \y & \y & \y & \y & \y & \n & \n 
& \n & \n & \n
& AMPAS, ANTLR-PD, ANTLR-PD-fallback \\

\rowcolor{black!6}
99 (11.7\%)
& \y & \n & \y & \y & \y & \y & \nul & \y & \n & \n
& \n & \n & \n
& Adobe-Display-PostScript, ADSL, Advanced-Cryptics-Dictionary \\

51 (6.0\%)
& \y & \n & \y & \y & \y & \y & \nul & \y & \n & \n
& \n & \y & \n
& 3D-Slicer-1.0, AdaCore-doc, Afmparse \\

\rowcolor{black!6}
48 (5.7\%)
& \y & \n & \y & \y & \y & \y & \n & \y & \n & \n
& \n & \n & \n
& AAL, Apache-1.0, Apache-1.1 \\

33 (3.9\%)
& \n & \n & \y & \y & \y & \y & \y & \y & \n & \n
& \n & \n & \n
& 0BSD, bcrypt-Solar-Designer, blessing \\

\rowcolor{black!6}
33 (3.9\%)
& \y & \y & \y & \y & \y & \y & \n & \y & \y & \y
& \y & \y & \n
& APSL-1.0, APSL-1.2, CATOSL-1.1 \\

29 (3.4\%)
& \y & \n & \y & \y & \n & \n & \n & \y & \n & \n
& \n & \n & \n
& Adobe-License, AdobeResearch, ASWF-Digital-Assets-1.0 \\

\rowcolor{black!6}
26 (3.1\%)
& \y & \n & \y & \y & \y & \y & \y & \y & \n & \n
& \n & \y & \n
& bzip2-1.0.5, bzip2-1.0.6, Catharon \\

23 (2.7\%)
& \y & \n & \y & \y & \y & \y & \n & \y & \n & \n
& \n & \y & \n
& CC-BY-3.0, CC-BY-3.0-AT, CC-BY-3.0-AU \\

\rowcolor{black!6}
21 (2.5\%)
& \y & \y & \y & \y & \n & \n & \n & \y & \n & \y
& \n & \n & \n
& BUSL-1.1, CC-BY-NC-SA-1.0, CC-BY-NC-SA-2.0 \\

\bottomrule
\end{tabular}}

\caption{
Top 10 most common license feature combinations in the SPDX corpus.
The first column reports the number of licenses exhibiting each feature
combination and its proportion of the corpus. Columns Attr--CL correspond to
the ten features in Nordlander's taxonomy \citep{nordlander2004}; Pat, StCh,
and Net denote the three extensions introduced by Kapitsaki
\citep{kapitsaki_open_source}. The final column lists representative licenses
for each feature profile. A $\checkmark$ indicates that a feature is present,
$\times$ indicates that it is absent, and $\cdot$ indicates that the property is
undefined or cannot be determined.
}

\label{tab:top_profiles}
\end{table*}
    
\section{Grading and measuring license (in)compatibility}

License selection is one application of the permissiveness framework introduced in this paper. While we attempt to classify permissiveness in this work, interpretations of permissiveness often depend on context, including: 1) developers’ own openness preferences, 2) the artifacts from which their work is derived, 3) the artifacts that may be derived from their technology, and 4) the technologies with which their system interacts. We showcase different methods throughout this work based on these contexts and suggest that the relational, comparative nature of pairwise preferences strikes an appropriate balance between capturing the tacit decision-making heuristics developers employ and providing the explicitness required for a generalizable framework.

 To be able to gauge the permissiveness of a license relative to others in the list, we would generally need to redo all pairwise comparisons. However, given the Bradley--Terry model we can use our pairwise orderings and transform them into a total order, where all the licenses can be assigned a latent score (the degree of permissiveness) driven by how much they are preferred across comparisons. We can then produce a ranked list of licenses based on these scores, where the probability that one license is more permissive than another, \(P(i \succ j)\), is calculated as:

\begin{equation}
  s(\ell) = \frac{\sum_{f \in F_\ell} w_f}{\sum_{f \in F} w_f} \in [0,1],
  \qquad
  \Delta = s(\ell_{\text{downstream}}) - s(\ell_{\text{upstream}}),
\end{equation}

where $F$ represents the set of directional features (each suggests permissiveness or restrictiveness), $F_\ell$ the subset that applies to a license $\ell$, and $w_f$ represents the weight of a feature's directionality.

Then, we use the Bradley--Terry model on our pairwise data to give us a total order to reference, which forms the basis of a lookup table, where \(i\) and \(j\) index two distinct licenses.
\[
\hat{P}
=
\bigl( P(i \succ j) \bigr)_{1 \le i,j \le n}
\]

This model helps us order new licenses without having to compare them against every previously compared license \citep{bradley_terry}. One application of this total order is a ranked list of licenses that serves as the backbone of a tool to filter licenses based on developer preferences. We define the permissiveness rank as follows. For a given license $L$, let $|L|$ denote the number of ecosystem licenses it is strictly more permissive than, and severity is $$\delta = \text{rank}(L_{\text{downstream}}) - \text{rank}(L_{\text{upstream}})$$ In the context of a developer choosing a license based on the degree of burden on a downstream developer and the number of licenses they can select, you can use this total ordering to produce a decision tree; where each node represents a feature, which in this case corresponds to a restriction. Starting with the root node, which if selected alone filters out the fewest licenses from our list, a developer can traverse the tree and add restrictions. As you go down the tree, the number of licenses available generally decreases as restrictions are applied. From this set of licenses, a developer can then choose any one of them and see its rank based on how many licenses below it are more restrictive. We determine the order of features in the tree based on their permissiveness, as determined by the linear regressions in Table~\ref{tab:nordlander-cascade}.

\begin{table}[tb]
  \centering
  \footnotesize
  \setlength{\tabcolsep}{3pt}
  \renewcommand{\arraystretch}{1.1}
  \resizebox{0.52\columnwidth}{!}{%
  \begin{tabular}{lrrr}
    \toprule
    Feature & $\beta$ & Remaining & Lost \\
    \midrule
    All licenses & --- & 843 & --- \\
    \midrule
    Allows commercial redistribution? & $+10.67$ & 661 & 182 \\
    \rowcolor{black!6}
    Allows commercial use of derivatives? & $+5.52$ & 646 & 15 \\
    No attribution required? & $+3.51$ & 55 & 591 \\
    \rowcolor{black!6}
    Allows modification? & $+2.48$ & 55 & 0 \\
    Allows derivative works? & $+2.48$ & 55 & 0 \\
    \rowcolor{black!6}
    No ShareAlike required? & $+2.21$ & 52 & 3 \\
    GPL-compatible? & $+1.68$ & 35 & 17 \\
    \rowcolor{black!6}
    No source disclosure required? & $+0.90$ & 35 & 0 \\
    Grants patent rights? & $+0.34$ & 0 & 35 \\
    \rowcolor{black!6}
    No state-changes obligation? & $+0.31$ & 0 & 0 \\
    Network use is not distribution? & $-0.05$ & 0 & 0 \\
    \rowcolor{black!6}
    Allows binary distribution? & $-0.51$ & 0 & 0 \\
    No copyleft? & $-1.11$ & 0 & 0 \\
    \bottomrule
  \end{tabular}}
  \caption{License features as preferences, with the number of licenses meeting each requirement. Features are drawn from the Nordlander and Kapitsaki taxonomies. The correlation between a feature and restrictiveness ($\beta$) typically narrows the set of licenses available to a downstream developer. The second column shows the latent factor for each feature, the middle column shows the number of licenses remaining given a chosen term, and the rightmost column shows licenses lost from the total pool.}
  \label{tab:nordlander-cascade}
\end{table}

We question whether we can use these scores to reconstruct the rankings from a feature-first perspective, the core benefit being that re-running pairwise judgments is an expensive task that scales quadratically ($O(n^2)$). We find that this is not the case: reproducing the order from feature-predicted scores $r_i^{\mathrm{BT}}$ rather than pairwise judgments, we see that 20.7\% of pairwise directions get reversed, with Spearman correlation ($\rho =  0.695$). We view this as motivation for why introducing new licenses into the ecosystem without considering their marginal legal contribution is a costly task both computationally and socially. Despite the ease of writing a new license, we can observe the actual costs to obtain additional model-based partial orderings, where pairwise judgments cost roughly $560\times$ more than deriving a list from extracted features ($\binom{n}{2}$ license pairs per model, compared with $n$ feature extractions). We describe the methodology and the loss in more depth in the appendix (Section \ref{sec:bt-total-ordering}, \ref{tab:rank-error-family}). We attribute loss to features being lossy representations themselves, where for all the pairs that were deemed incomparable, 16.3\% of model rationale pointed to features that were not in the library we employ (the average of the models: GPT-5.5, 17.5\%; Opus 4.6 16.1\%; Qwen 3.5 15.2\%). When considering BT on pairwise verdicts, BT had a contradiction rate that was 0.7\% on the pairs where all LLM raters agreed. 


A developer can consider license choice as a set of preferred restrictions, where for every combination of chosen restrictions a a hypothetical license exists. In practice, the lack of such a license would suggest that a license with that combination could be seen as \emph{functionally distinct} compared to the existing ones in the list. This tool allows a developer to understand how the decisions they care about affect what licenses are available to them and how this affects developers who build on top of their technologies.  Many licenses have similar combinations as observed in Table \ref{tab:top_profiles}, and these licenses can be interpreted as functionally equivalent based off the features we investigate. 

Agreement is employed to understand the validity of the rankings derived from features compared to more expensive pairwise orderings. We find that Spearman correlation demonstrates that BT cannot successfully derive a reliable order given features alone. However, grouped agreement is meaningfully higher (with a 20 point increase). While features could be a means to understanding ecosystem-level patterns efficiently (f.ex. understanding the degree of permissiveness across a corpus or seeing the most popular combinations), feature-first analysis appear best for studying local instead of global structure.

\section{Case studies}
Previous work has explored incompatible license choices for models on 
Hugging Face, where downstream models (i.e.\ fine-tuned, quantizations, 
model merges, and adapters) are accompanied with more permissive licenses than their parents, but to our knowledge, these works lacked a framework that could formally classify this phenomenon or the extent of its implications
\citep{laufer2025anatomymachinelearningecosystem}. We take these 
observations as motivation and study a snapshot of five popular 
software library ecosystems, covering 9.3M+ packages: npm ($\sim$8.2M), Maven ($\sim$546K), PyPI ($\sim$444K), Cargo ($\sim$139K), and conda-forge ($\sim$32K), from 2010 to 2025.\footnote{Sourced from a snapshot from March 30, 2026}

A question a developer or legal expert might be interested in is the degree of infringement by choosing an incompatible license. Not all incompatibilities are the same, but how do you compare them? We consolidate the feature-wise and partial order findings to grade incompatibilities. For a given tree of (software library) dependencies (Figure~\ref{fig:dep-tree-example}), you can see the degree of violation for one dependency to its parent or take the sum of violations to understand how they propagate through a tree.

Where the former is important for some developer selecting a license and the latter is more important for an enforcer of licenses who cares to `measure' infringement. We consider the library ecosystems and use these methods to compare them. 

 While traversing a tree, we take every adjacent license in some lineage (depth n, and depth n-1) and perform the same operation. With all of these deltas, we take the sum of the positive values to get a score of incompatibility for the tree.

As an illustrative example, we can consider the following tree with descendants of the \texttt{requests-oauthlib} PyPI package---used for adding authentication support in Python apps---of depth 2 (indexed at 0). At depth 0, there's the \texttt{requests-oauthlib} package with an ISC license; at depth 1, there's the \texttt{oauthlib} package, which is lower-level, with a BSD-3-Clause license, and \texttt{requests} with an Apache-2.0 license. It is at depth 2 where the biggest incompatibility occurs, between \texttt{requests} with an Apache-2.0 license and \texttt{certifi} with an MPL-2.0 license. Their permissiveness scores differ by 0.09. By their profiles, we see that there are 1,130 ordered pairs of licenses with the same degree and type of incompatibility divergence (a function of the features that differ).

  \begin{figure}[H]
    \centering
    \resizebox{0.9\columnwidth}{!}{%
    \begin{tikzpicture}[
      grow=right,
      every node/.style={font=\ttfamily\scriptsize, anchor=west},
      edge from parent/.style={draw, ->, >=stealth, thick,
        edge from parent path={(\tikzparentnode.east) -- (\tikzchildnode.west)}},
      level distance=4.6cm,
      level 1/.style={sibling distance=2.2cm},
      level 2/.style={sibling distance=1.05cm},
      lab/.style={font=\sffamily\tiny, midway, above, sloped, inner sep=1.5pt},
    ]
    \node {requests-oauthlib {\normalfont\tiny(ISC)}}
      child {node {requests {\normalfont\tiny(Apache-2.0)}}
        child {node {charset-normalizer {\normalfont\tiny(MIT)}}
          edge from parent node[lab] {$\Delta = -0.06$}}
        child {node {idna {\normalfont\tiny(BSD-3-Clause)}}
          edge from parent node[lab] {$\Delta = -0.06$}}
        child {node {urllib3 {\normalfont\tiny(MIT)}}
          edge from parent node[lab] {$\Delta = -0.06$}}
        child {node[text=red!80!black] {certifi {\normalfont\tiny(MPL-2.0)}}
          edge from parent node[lab] {$\Delta = +0.09$}}
        edge from parent node[lab] {$\Delta = +0.06$}}
      child {node {oauthlib {\normalfont\tiny(BSD-3-Clause)}}
        edge from parent node[lab] {$\Delta = 0$}};
    \end{tikzpicture}}
    \caption{Example software library dependency tree in PyPI. Each node is a package and its declared license, and each edge carries $\Delta$. A violation occurs when a downstream package carries a more permissive license than a dependency: \texttt{requests} (Apache-2.0) depends on \texttt{certifi} (MPL-2.0), a more restrictive license, whereas its other three dependencies are more permissive than it.}
    \label{fig:dep-tree-example}
  \end{figure}
  
  Here, we set out to investigate similar patterns in the open-source software ecosystem. To do this, we first need to map relationships between software packages. We use deps.dev---a Google service which has dependency graphs for software packages, we then classify the edges representing dependencies from a snapshot of the dataset \citep{depsOpenSource}. We only look at licenses that exist within a chain and do not look at isolated packages. We define consistency as whether a downstream artifact's license is equally or more restrictive than its parent. Cargo ($\sim$83\%) has the highest ratio of consistent license relations, followed by PyPI ($\sim$54\%), Maven ($\sim$53\%), then npm ($\sim$48\%) and conda-forge ($\sim$41\%) (Figure \ref{fig:severity-violations}). Grading allows us to understand incompatibility from an ecosystem-level perspective. When looking at the weight of the violations in an ecosystem and not only how many, the ranking re-orders, and the highest infringement pairs of dependencies per ecosystem are as follows. The most common violations are also common across ecosystems (MIT --> Apache 2.0). Ecosystem-wise, we can try and diagnose why this is the case. Cargo's build-time checks for imports, the libraries that the given repository depends on but not the license that a developer might choose for their work.

\begin{table}[H]
    \centering\footnotesize
    \setlength{\tabcolsep}{4pt}
    \begin{tabular}{llrrl}
      \toprule
      Ecosystem & Most common violating pair & $\Delta$ & Share & Pair carrying most summed $\Delta$ \\
      \midrule
      PyPI  & MIT on PSF-2.0    & 0.07 & 36\% & the same, 31\% \\
      \rowcolor{black!6}
      Maven & MIT on Apache-2.0 & 0.06 & 68\% & the same, 48\% \\
      Cargo & MIT on Apache-2.0 & 0.06 & 36\% & CC0 on MIT (0.19), 32\% \\
      \rowcolor{black!6}
      npm   & MIT on Apache-2.0 & 0.06 & 50\% & CC0 on MIT (0.19), 40\% \\
      \bottomrule
    \end{tabular}
    \caption{Most common violating license pairs per ecosystem, downstream on upstream, with their share of the ecosystem's violating edges and the pair contributing most to the summed $\Delta$.}
    \label{tab:violating-pairs}
\end{table}

  \begin{figure}
      \centering
      \includegraphics[width=0.5\linewidth]{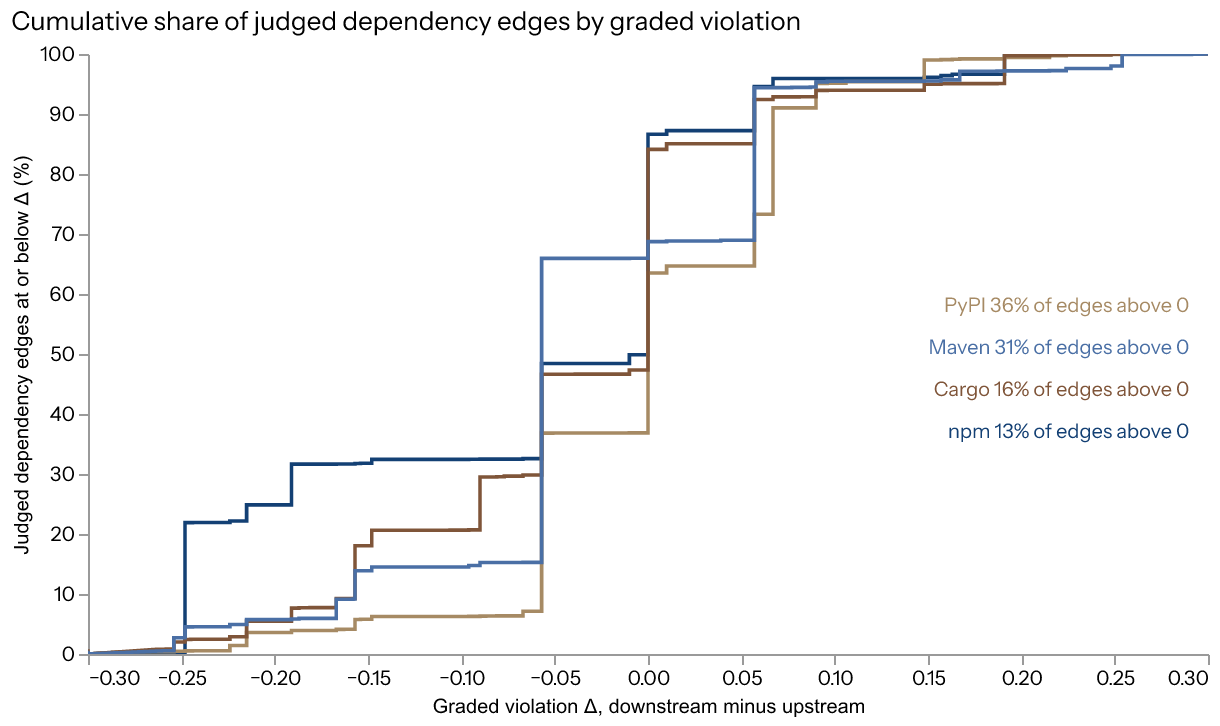}
       \caption{Distribution of incompatibilities across library ecosystems, where the shape of the distributions are similar across all ecosystems. PyPI has the largest share of incompatible dependencies, but Cargo has the highest severity per incompatibility.}
      \label{fig:incompatibility-distribution}
  \end{figure}

\section{Limitations}

While we rely on existing, bottom-up license taxonomies to inform the features we look into, the space of possible features is large. For example, we broadly refer to license terms as features, but legal scholars have explored the different types of legal relations encoded in licenses, and a more granular treatment of these obligations and restrictions remains an avenue for future work \cite{hohfeld1913}. We see the choice of features in this paper more as a case study into what is possible, and expect that the method can apply to a wider range of features, especially in domain-specific settings.

\subsection{Understanding incomparability}

The ML model ecosystem is a breeding ground to study license relationships due to the composite nature of technologies (e.g., fine-tuned or distilled models, training data, and generative outputs). In particular, our work investigates \textit{comparability} between licenses, which we find is not always equivalent to \textit{compatibility} along license chains. For instance, a downstream license may be more permissive than an upstream license, which may not necessarily violate the upstream license restrictions. The reverse is also true: a downstream license may be more restrictive than an upstream copyleft license, but violates the share-alike requirements for derivative licenses.
 
 
 Furthermore, licenses themselves get iterated upon: consider if a license is an earlier version of the same license---which terms should apply? Our exploration surfaces 55 cases where a model has one recognized license selected in its platform metadata but a different license listed in the model card’s \texttt{license\_name} field. 
 
Furthermore, the definitions of the model relationships is far from concrete. Clarifying these ambiguities will be key for formalizing automated legal processing approaches. Here, we consider the lack of a feature or clause as suggesting that it is false, but it could be interpreted as nullification. Indeed, models are prone to amplifying gaps in human reasoning but the criticality of the domain motivates its exploration.

\section{Conclusion}

In this work, we consider whether LLMs can be used to recover global structure where local structure is hard to define. While it is possible to look into the substructure of licenses, analyzing the clauses themselves, we focus on interpreting the network of licenses themselves in a top-down manner to see if reliability can be gleaned beyond model-defined confidence which is typically unreliable. And we apply this framework to software licenses where the implications are timely and pressing. We use pairwise comparisons to study the networks of software licenses and apply this partial ordering to existing software ecosystems.


\section*{Acknowledgements}

We thank James Grimmelmann, Jon Kleinberg, and Anna Kazlauskas for their feedback as well as Taste Labs and Open Data Labs for their in-kind support.

\bibliographystyle{acl_natbib}
\bibliography{references}

\clearpage
\appendix

\section{Appendix}

\subsection{Total ordering from extracted features}
\label{sec:bt-total-ordering}

\begin{table}[h]
  \centering
  \footnotesize
  \setlength{\tabcolsep}{5pt}
  \renewcommand{\arraystretch}{1.1}
  \begin{tabular}{lrrrr}
    \toprule
     & Agree & Disagree & Tie & Agree where non-zero \\
    \midrule
    Fitted strengths, all verdict pairs        & 74.5\% & 14.2\% & 11.4\% & 84.0\% \\
    \rowcolor{black!6}
    Fitted strengths, held out by license      & 74.0\% & 14.3\% & 11.6\% & 83.8\% \\
    Equal weights, all verdict pairs           & 70.5\% & 12.2\% & 17.3\% & 85.2\% \\
    \rowcolor{black!6}
    Fitted strengths, judged dependency edges  & 65.6\% &  0.1\% & 34.3\% & 99.8\% \\
    Equal weights, judged dependency edges     & 38.6\% &  0.1\% & 61.2\% & 99.7\% \\
    \bottomrule
  \end{tabular}
  \caption{Pairwise ordering agreement by feature-based representation. Held out by license means no license in a scored pair was seen while fitting.}
  \label{tab:graded-agreement}
\end{table}

More formally, we can describe the rank recovery process as the following. First, we fit the feature-based rankings and the BT (ordering) model. Feature-based rankings are derived from the scores taking the sum of the learned latent factors. We compare the derived ranking from the ground truth partial order by looking at all the directed pairs ($i, j$) where i $i \succ j$ or $j \succ i$. To compare the full lists in terms of similarity which offers a more holistic view than agreement alone, we also look at Spearman correlation and extend it to consider the fact that because we have a partial order we have groups of comparable licenses. Consider a number of ordered lists, where every item is a comparable license of a $a > b > c > d$, we first count the traditional agreement across pairs and then the average of these agreements across groups. 

\begin{enumerate}
    \item The first measure $\rho$ is defined as:
    \[
    \small
    \rho
    =
    \frac{\text{\# agreeing directed pairs}}{\text{\# total directed pairs}}
    =
    \frac{C}{C + D} = 0.54
    \]

    \item For agreement across $g$ grouped ordered lists of licenses
    $\{\mathcal{L}_1,\ldots,\mathcal{L}_g\}$,
    let $C_i$ and $D_i$ denote the number of agreeing and disagreeing
    pairwise-ordered pairs within group $\mathcal{L}_i$. If we ever
    report a the average of the agreement across groups, which we define as:
    \[
    \rho_{\mathrm{grouped}}
    \;=\;
    \frac{1}{g}\sum_{i=1}^{g}\frac{C_i}{C_i + D_i} = 0.74
    \] 
\end{enumerate}

\subsubsection{Bradley-Terry model loss}
\begin{table}[h]
  \centering
  \footnotesize
  \setlength{\tabcolsep}{5pt}
  \renewcommand{\arraystretch}{1.1}
  \begin{tabular}{lrr}
    \toprule
    \textbf{Family} & $n$ & \textbf{Mean rank error} \\
    \midrule
    Permissive & 18 & 2.0 \\
    Data       & 19 & 2.5 \\
    Copyleft   & 35 & 3.3 \\
    ML         & 19 & 3.9 \\
    \bottomrule
  \end{tabular}
  \caption{Rank error by license family.}
  \label{tab:rank-error-family}
\end{table}

\subsection{Additional figures}

\begin{figure}
    \centering
    \includegraphics[width=0.55\linewidth]{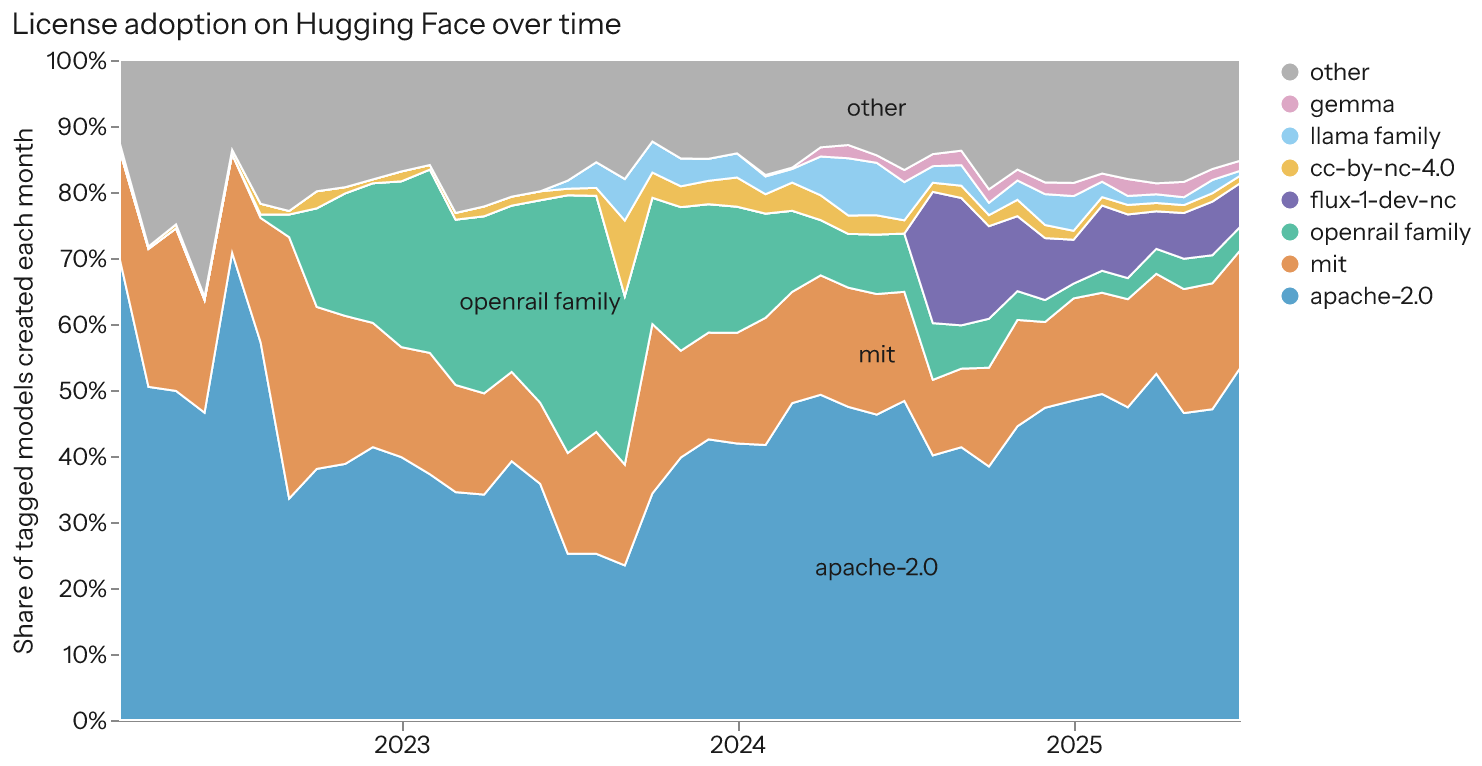}
    \caption{Adoption of licenses on Hugging Face}
    \label{fig:license_model_share_hf}
\end{figure}
\begin{figure}[tb]
    \centering

    \begin{subfigure}[t]{0.5\linewidth}
        \centering
        \includegraphics[width=\linewidth]{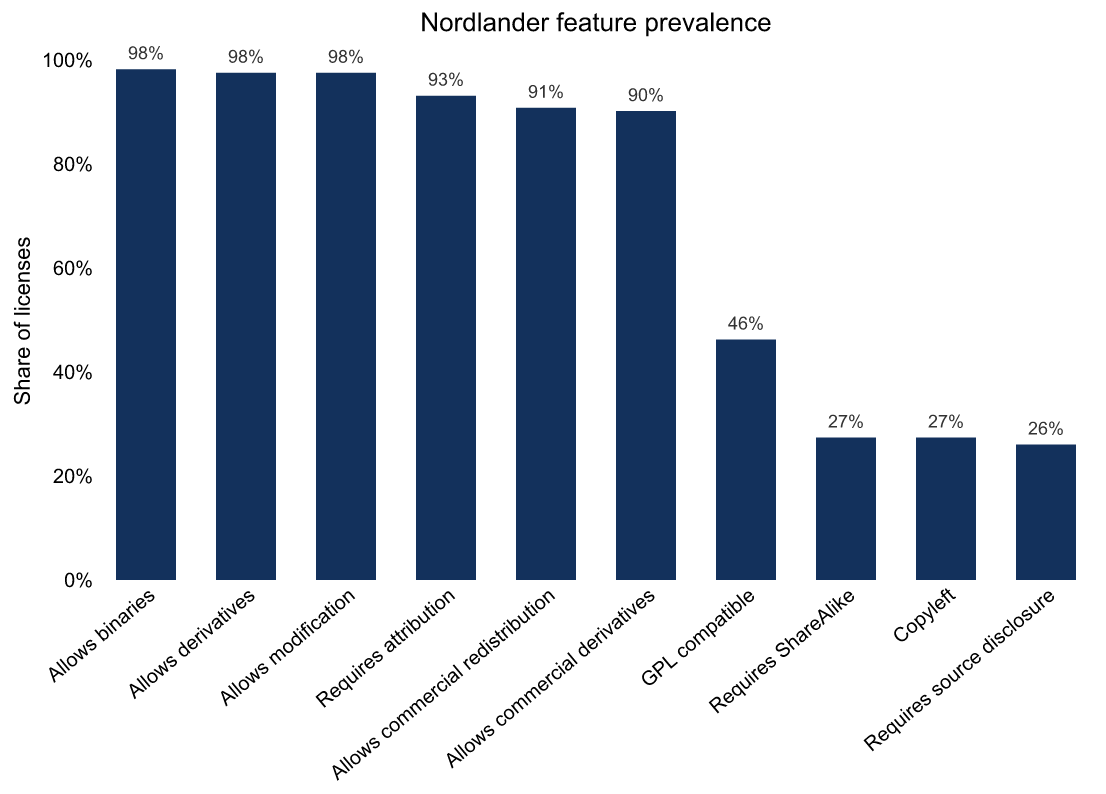}
        \caption{Nordlander features}
        \label{fig:nordlander-prevalence}
    \end{subfigure}
    \hfill
    \begin{subfigure}[t]{0.3\linewidth}
        \centering
        \includegraphics[width=\linewidth]{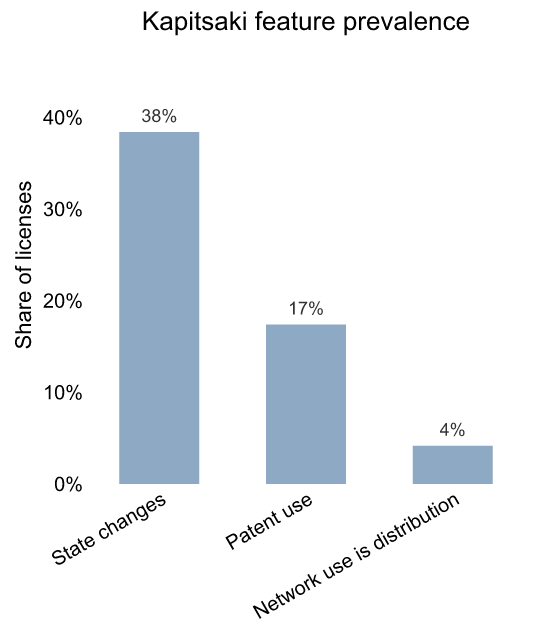}
        \caption{Kapitsaki features}
        \label{fig:kapitsaki-prevalence}
    \end{subfigure}

    \caption{Feature prevalence across taxonomies. }
    \label{fig:feature-prevalence}
\end{figure}

\begin{figure}[H]
    \centering
    \includegraphics[width=0.62\linewidth]{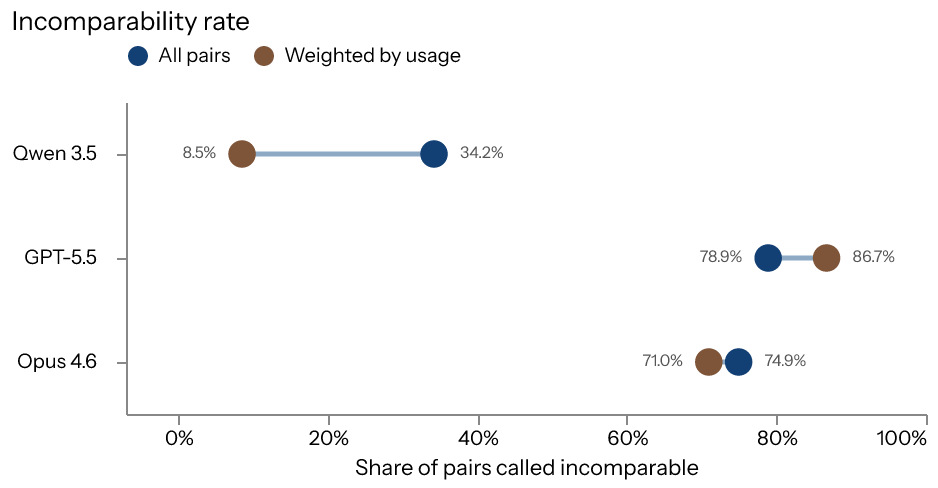}
    \caption{Incomparability rates by license popularity.}
    \label{fig:licenses_popularity_by_incomporability}
\end{figure}

\begin{figure}
    \centering
    \includegraphics[width=0.72\linewidth]{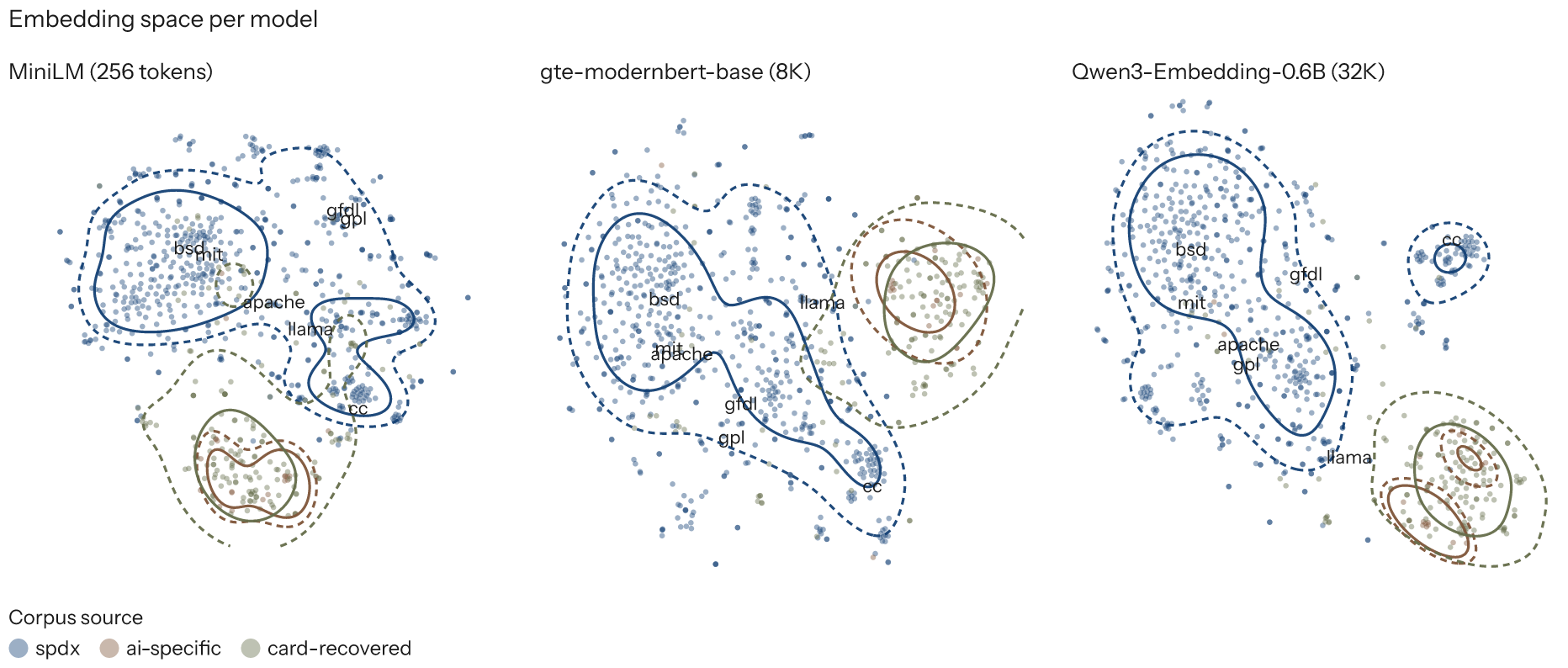}
    \caption{License embedding atlas showing the discrimination of the space by family with t-SNE as the dimensionality reduction algorithm. Counters are outlined to show the cluster density at 80\% and 50\% concentration.}
    \label{fig:embedding-atlas}
\end{figure}

\begin{figure}[tb]
    \centering
    \includegraphics[width=0.58\linewidth]{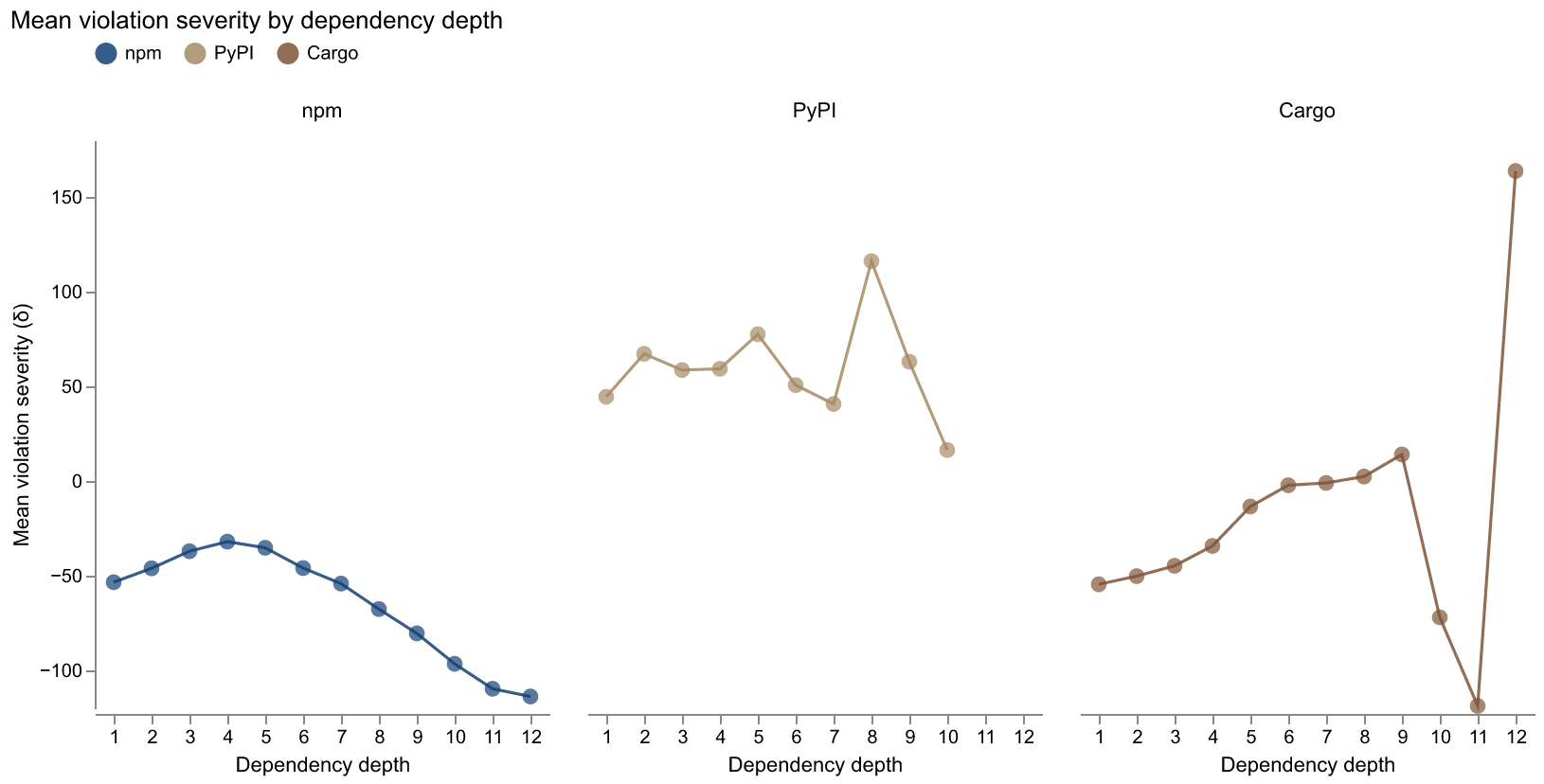}
    \caption{Severity of license inconsistencies (violations of permissive ordering) by software package depth.}
    \label{fig:severity-violations}
\end{figure}
 Cargo's high consistency may be attributed to the tools the community
  gives its developers, \texttt{cargo-deny} and \texttt{cargo-about} give
  software maintainers \emph{executable governance} at build time to check
  if the dependencies they have chosen for a project abide by the terms of
  the license they choose. More formal license fields are promoted for use
  in \texttt{Cargo.toml} files which are structured and machine-readable
  and can be seen as programmatic documentation that can be more easily
  verified. 

\subsection{Additional related work}

Additional previous work has framed open-source licenses as infrastructure for governing the technological commons \citep{ChoksiGrimmelmann2024HowLicensesLearn, Ostrom2010Beyond}, where each license reflects the values of its inventors and their design constitutes collective deliberation over the needs of downstream developers. One reason why the optimal licensing regime is hard to conceive is that technologies are built in tandem and on top of each other within a broader supply chain; understanding what rules apply to whom, especially when they are often vaguely specified, is a difficult task. 
\subsection{Limitations}

\subsubsection{License dataset}

To the best of our ability, we note and acquire a list of licenses and 
their text, but recognize that the diffuse nature of license authorship means that there are likely licenses on smaller platforms that may have been missed. 

\subsubsection{Ordering bias}
Given a pair of licenses to be compared, it is possible that a
  model would have a tendency to assign a judgment based on the order in which
  they are presented. We explored this independently on a small sample and find
  that there is 76\% consistency on a set of 50 randomly selected license pairs,
  where 12 out of 50 verdicts flipped when the order of licenses flipped. We find
  that the majority of the flips favored the license that appeared first. Of the
  12 flips, 11 were marked as incomparable, suggesting that when the model is less
  confident this bias is a stronger influence, and that when a model reasons and
  commits to a confident verdict, positional bias is less of a factor. Future work
  could explore these tendencies in more depth (e.g.\ at a larger scale with
  order-balanced presentation).

\subsubsection{Software libraries as a case-study}

As part of our case-studies, we look to the software package ecosystem as a way to apply this work 
but note that there are shortcomings of processing data at scale \cite{wuempiricalopenstudylicenses}. We 
assume that all SPDX identifiers are unique and that all packages have 
a corresponding license where applicable, but realize that if they are 
not, there may be duplications and sparsity in the dataset. 
The \texttt{deps.dev} \texttt{MinimumDepth} parameter 
represents the shortest path from a library to a dependency across all 
dependency trees, but we acknowledge that it does not show the full 
depth of the tree.

 \begin{table}[tb]
    \centering 
    \begin{tabular}{lrrr}
    \toprule
    Ecosystem & Max depth & p50 & p99 \\
    \midrule
    npm         & 199 & 4 & 12 \\
    PyPI        &  17 & 2 &  5 \\
    Cargo       &  14 & 3 &  7 \\
    Maven       &  15 & 2 &  7 \\
    conda-forge &   6 & 1 &  2 \\
    \bottomrule
    \end{tabular}
    \caption{Length of compatible dependency chains across software packages.}
    \label{tab:depth_distribution}
  \end{table}

\begin{table}[tb]
    \centering\footnotesize
    \resizebox{0.52\columnwidth}{!}{%
    \begin{tabular}{lrrrr}
      \toprule
      Ecosystem & Compliant & Incomparable & Violation & $n$ edges \\
      \midrule
      npm         & 47.7\% & 39.8\% & 12.6\% & 2.5B \\
      PyPI        & 54.2\% &  3.8\% & 42.0\% & 53.1M \\
      Cargo       & 83.1\% &  4.6\% & 12.3\% & 30.5M \\
      conda-forge & 40.9\% &  1.6\% & 57.4\% & 122K \\
      Maven       & 53.1\% &  0.0\% & 46.9\% & 66.2M \\
      \bottomrule
    \end{tabular}}
    \caption{License relations across software package ecosystems. Incomparability is relatively low across all packages ($\leq$5\%), with the exception of npm ($\sim$40\%).}
    \label{tab:verdict_by_ecosystem}
  \end{table}

\subsection{Prompt iterations}

We iterated on prompts in later runs, but the plots in the main paper use an earlier version without an explicit definition of logical entailment; we compare prompt results and inter-model agreement below. We run the most recent prompt version with a panel of models (GPT 5.5, Opus 4.6, Qwen 3.5) which we select for their similar release dates and strong performance on legal reasoning tasks while being from performant closed and open model families \cite{vals_ai}. We use these models on a smaller, selectable subset of Hugging Face licenses (93 licenses) which represent the majority of license selection. Improvements include removing model self-confidence as an unnecessary measure, a simplified output schema, instructing models to assess incompatibility as a separate, additional step from determining the main verdict, and an enhanced definition clarifying that incompatibility refers to restrictions operating on distinct dimensions (or those that are ``orthogonal'' to one another). While we believe these improvements are justified, the plots and findings in the main paper reflect earlier iterations; we find the reported findings stay consistent on the HF set and hope the improved versions can be used and shared publicly for future analysis.

\begin{table}[tb]
    \centering
    \footnotesize
    \begin{tabular}{llrrr}
      \toprule
      Model pair & & $n$ & Agreement & $\kappa$ \\
      \midrule
      GPT-5.5 vs Opus 4.6  & orderings & 4{,}276 & 94.2\% & 0.883 \\
      \rowcolor{black!6}
      GPT-5.5 vs Qwen 3.5  & orderings & 4{,}270 & 91.3\% & 0.828 \\
      Opus 4.6 vs Qwen 3.5 & orderings & 4{,}270 & 91.3\% & 0.828 \\
      \midrule
      \rowcolor{black!6}
  GPT-5.5 vs Opus 4.6  & features & 10{,}777 & 94.2\% & 0.888 \\
      GPT-5.5 vs Qwen 3.5  & features & 10{,}777 & 95.7\% & 0.915 \\
      \rowcolor{black!6}
      Opus 4.6 vs Qwen 3.5 & features & 10{,}777 & 93.9\% & 0.883 \\
      \bottomrule
    \end{tabular}
    \caption{Agreement between models for pairwise verdicts on the HF-93 sample (93 licenses selectable on Hugging Face).}
    \label{tab:panel-agreement}
\end{table}

\subsection{Prompts used for LLM analysis}

\begin{promptbox}{Pairwise ordering prompt}

\textbf{Task.} Given two license texts, determine their relative permissiveness.

\textbf{Step 1. Ordering (required).}

\begin{itemize}\itemsep0pt\parskip0pt\topsep2pt
  \item \texttt{A > B}: License A is strictly more permissive than B.
  \item \texttt{A = B}: Licenses A and B have equivalent scope, with the same
        net permissions and obligations, even if expressed using different
        wording (e.g., MIT and ISC, or CC0 and the Unlicense).
  \item \texttt{A < B}: License B is strictly more permissive than A.
\end{itemize}

An equivalence judgment should only be returned when the licenses have the same
substantive permissions and obligations; it should not be used as an abstention.

\textbf{Step 2. Incomparability.}

Set \texttt{incomparable=true} only when the licenses impose a tradeoff across
distinct dimensions: one license is more permissive on at least one dimension
while the other is more permissive on another. Set
\texttt{incomparable=false} when the licenses have a clear ordering or are
equivalent.

\textbf{Output schema.}

\begin{verbatim}
{
  "ordering_verdict":
      "A > B | A = B | A < B",
  "incomparable": true | false,
  "incomparability_summary":
      "<one or two sentences>" or null,
  "insufficient_information":
      true | false
}
\end{verbatim}

\end{promptbox}

\begin{promptbox}{Feature extraction prompt}

\textbf{System.}
Extract structured features from license texts. Return only a JSON object with
no explanation or additional text. Each feature should be assigned
\texttt{true}, \texttt{false}, or \texttt{null}. Use \texttt{null} only when the
license text does not provide sufficient information to determine the value.

\textbf{User.}
Answer the ten feature questions from the Loreto, Oliner \& Woo (2004)
software license taxonomy listed in Table~\ref{tab:license_features}. Use the
license text as the sole source of truth. Questions should be interpreted using
their original taxonomy definitions, for example
\texttt{requires\_attribution} corresponds to ``Copyright Notice Must Be
Attached?'' and \texttt{gpl\_compatible} corresponds to ``GPL Compatible?''.

Feature defaults are asymmetric by design. Attribution, modification, and
derivative-work permissions default to \texttt{true} unless explicitly
restricted. Commercial redistribution defaults to \texttt{true} when the text
is silent. \texttt{gpl\_compatible} should return \texttt{null} when
compatibility cannot be established from the license text alone.

The model receives the license name and license text and returns a flat JSON
object containing all ten fields.

\end{promptbox}

\begin{figure*}[t]
\begin{verbatim}
{
  "ordering_verdict": "A > B | A = B | A < B",
  "incomparability_analysis": {
    "verdict": "incomparable | comparable",
    "factors": [
      {
        "type": "<short label>",
        "explanation": "<one or two sentences>",
        "evidence_by_license": [
          {
            "license": "A",
            "quote": "<short exact excerpt from A>",
            "location": "<section or heading, or null>"
          },
          {
            "license": "B",
            "quote": "<short exact excerpt from B>",
            "location": "<section or heading, or null>"
          }
        ]
      }
    ]
  },
  "insufficient_information": {
    "verdict": true | false,
    "factors": [
      {
        "type": "<short label>",
        "explanation": "<one or two sentences>"
      }
    ]
  }
}
\end{verbatim}
\caption{Output schema for the v8 pairwise prompt. Each asserted incomparability
factor must be supported by a verbatim excerpt from both licenses. Factors
without textual evidence should not be reported.}
\label{fig:v8-schema}
\end{figure*}

\subsection{Model calibration}
We use multiple models to calibrate our confidence in LLM verdicts. To study coherence, we use a lookup table with every license pair and a corresponding verdict. To study whether there are contradictions, we look for cases where a model judges license $A$ as more permissive than $B$ in one instance but less permissive in another. We find no such contradictions. Moreover, we check for cycles, where $A > B$ and $B > C$ but $C > A$. Out of 129{,}766 groups of three votes from each model, we count cycles in 1\% from the HF-93 sample. In other words, we find little Condorcet cycles, indicating that the judgments are transitive and coherent, as they form a consistent ordering of verdicts \citep{arrow_kenneth_social_choice}. We also report the inter-rater agreement (between models) for the extracted binary features. We compute Cohen's $\kappa$, where each feature has three possible values (\texttt{True}/\texttt{False}/\texttt{null}), and find that the average agreement across models is $0.8952$ and head-to-head comparisons are as follows: $\kappa$ = 0.888 for GPT-5.5 against Opus~4.6, $0.915$ for GPT-5.5 against Qwen~3.5, and 0.883 for Opus~4.6 against Qwen~3.5, over 10,777 pairs. The anomaly is the GPL compatibility feature, where we see relatively low agreement. This is likely because it is a feature denoted in the license identifier itself instead of in the text.

 We also compute the Kemeny order, the total order that aligns with the greatest number of pairwise comparisons, and we see that 8 out of 3,757 verdicts where all three models agree are contradicted.

\begin{table}[tb]
  \centering\footnotesize
  \setlength{\tabcolsep}{4pt}
  \begin{tabular}{@{}lrrrrr@{}}
    \toprule
    Rater & Edges & Cyclic & \% & Largest & Incomp. \\
          &       & triples &   & SCC     & \%      \\
    \midrule
    GPT-5.5  & 4{,}252 &   777 & 0.60 & 67 & 78.9 \\
    \rowcolor{black!6}
    Opus 4.6 & 4{,}251 &   916 & 0.71 & 50 & 74.9 \\
    Qwen 3.5 & 4{,}221 & 2{,}083 & 1.61 & 87 & 34.2 \\
    \bottomrule
  \end{tabular}
  \caption{Condorcet cycles and logical coherence of pairwise orderings for the HF-93 license set. The largest structural connected component is enumerated in the third column.}
  \label{tab:coherence}
  \end{table}

\begin{figure}
    \centering
    \includegraphics[width=0.55\linewidth]{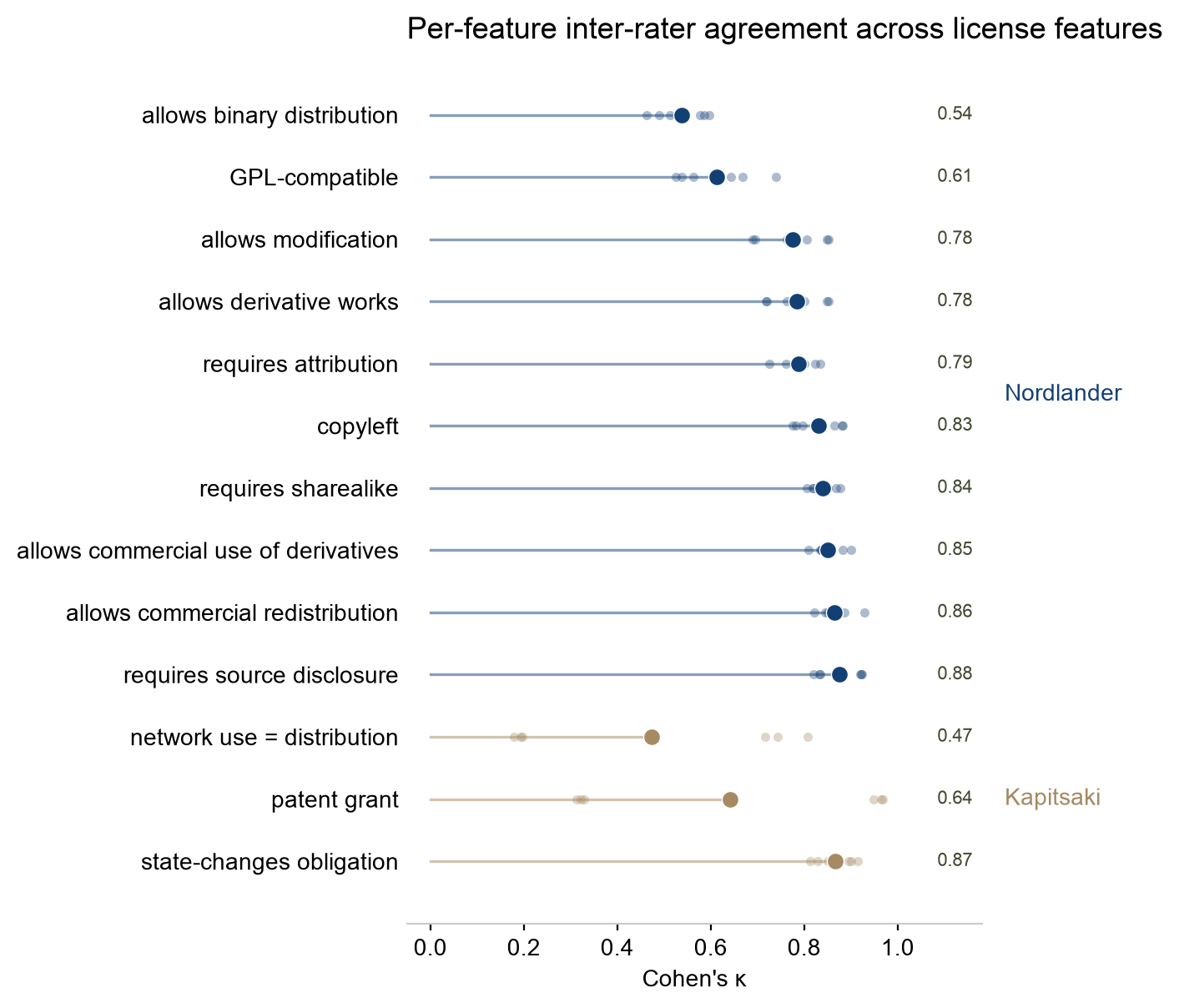}
    \caption{Cohen's $\kappa$ analysis for features between Sonnet 4.6 and DeepSeek. For all extractions, temperature was set to 0.}
    \label{fig:inter-rater-feature}
\end{figure}

\begin{figure}[tb]
    \centering
    \includegraphics[width=0.58\linewidth]{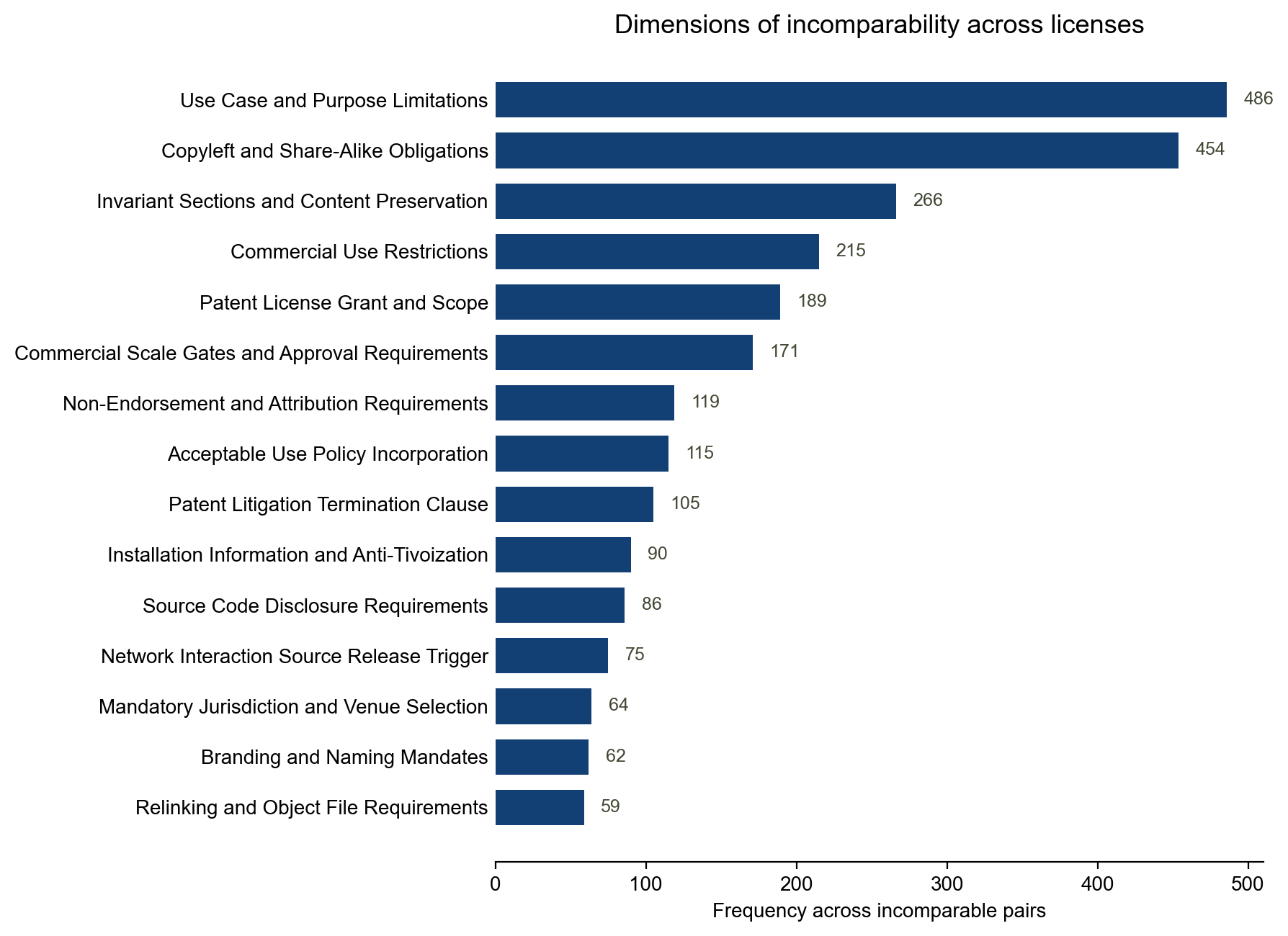}
    \caption{Breakdown of incomparability rationales. Haiku 4.5 is used for canonicalization phase of theme extraction due to its efficiency and the simpler task of classifying themes against a fixed list compared to initial parsing which requires deeper reasoning where Sonnet 4.6 is used.}
    \label{fig:incomparability_themes}
\end{figure}

In our prompt, we give models the ability to independently deem two licenses incomparable, instructing them to do so only after determining a verdict so that we can study the nature of incomparability itself. In earlier experiments, models were liberal in flagging incomparability when given the opportunity upfront, which motivated this order of operations. To extract the most common themes from model rationales, we prompt an LLM to identify core themes for each rationale separately, then canonicalize the resulting themes by passing the joint list to another model instance that groups semantically similar ones, which we then count. Below (Figure \ref{fig:incomparability_themes}), we plot the most common themes by frequency across incomparability rationales .We count 3,718 incomparable pairs out of 4,278 total ($\binom{93}{2}$ from the HF-93 license set) under the v7.1 prompt with Sonnet 4.6. We then plot the top 15 of 28 categories.

 \begin{promptbox}{Phase 1: Taxonomy derivation}
  You are analyzing descriptions of why pairs of software licenses are legally
  incomparable. Each description explains which specific legal dimensions put the
  two licenses in tension --- e.g., one imposes copyleft obligations the other does
  not, or one grants patent rights while the other restricts patent use. 
  
  Below are 80 representative descriptions:
  \{descriptions\}
  
  Task: Derive a canonical taxonomy of 20--28 categories covering the legal
  dimensions that make licenses incomparable. Each canonical category should:
  - Have a short, precise label (3--7 words, title case)
  - Represent a coherent, distinct legal dimension or obligation type
  - Be specific enough to be informative 
  
  Return ONLY a JSON array of canonical category label strings, nothing else.
  \end{promptbox}
  
  \vspace{0.5em}
  
  \begin{promptbox}{Phase 2: Mapping (Haiku~4.5)}
  Canonical legal dimension categories:
  \{taxonomy\}
  
  Each item below is a 1--2 sentence description of why two software licenses
  are incomparable. Map each to the single most relevant canonical category.
  Return ONLY a JSON object mapping the exact input string to the canonical
  category label. If none fit, use ``Other''. 
  
  Descriptions:
  \{descriptions\}
  \end{promptbox}


 \subsection{Human calibration}
 
 We employ a human panel from Prolific and internally (among authors) to annotate a sample of licenses for their features and pairwise orderings. 
 
 The following 16 licenses are used for annotation sample, with nine pairs drawn from the set: 3D-Slicer-1.0, AAL, ADSL, AFL-2.0, AFL-2.1, AGPL-1.0-or-later, AGPL-3.0-only, Aladdin, BSD-2-Clause-NetBSD, BSD-Attribution-HPND-disclaimer, BSD-Systemics, CDLA-Permissive-1.0, EUPL-1.0, Fair, GFDL-1.2-no-invariants-or-later, RSCPL. Raters are asked to revise the Nordlander features only. 

 \begin{figure}
     \centering
     \includegraphics[width=0.66\linewidth]{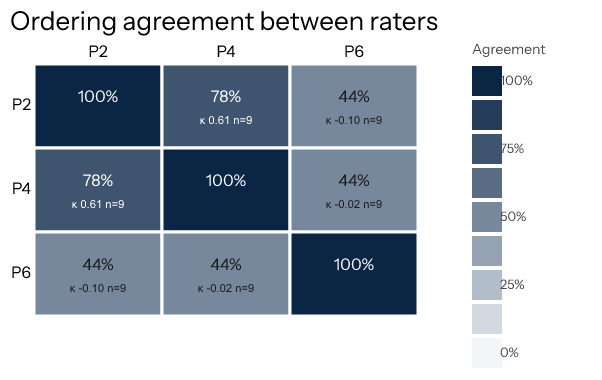}
     \caption{Inter-rater agreement for human annotators on pairwise orderings.}
     \label{fig:irr-pairwise-human}
 \end{figure}

  \begin{figure}
     \centering
     \includegraphics[width=0.52\linewidth]{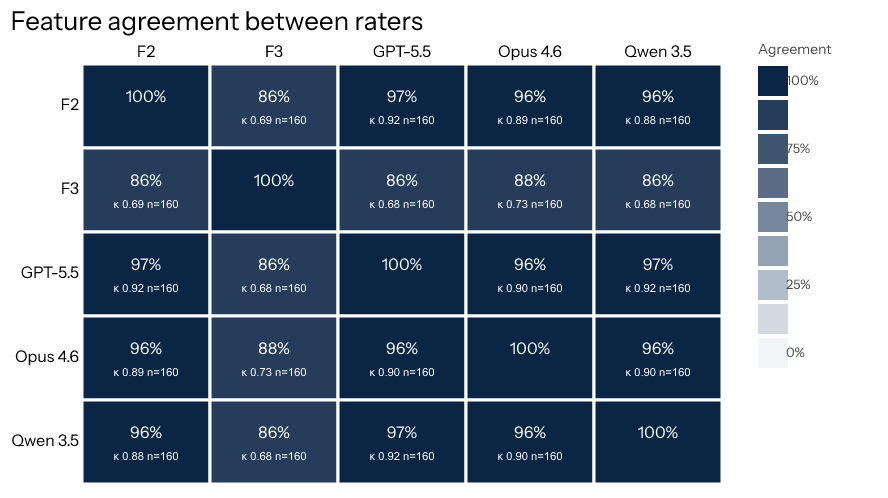}
     \caption{Inter-rater agreement for human and machine annotators on feature labels.}
     \label{fig:irr-feature-human-model}
 \end{figure}

 \begin{figure}
     \centering
     \includegraphics[width=0.62\linewidth]{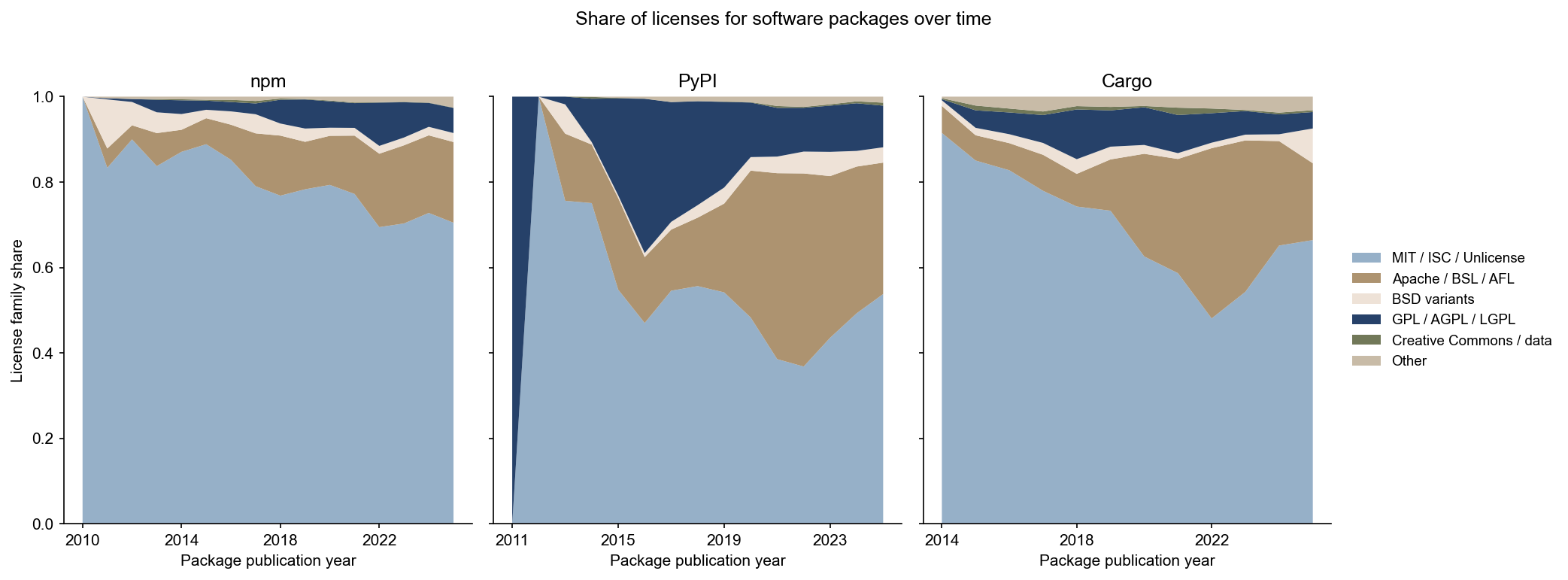}
     \caption{Share of licenses across three popular software ecosystems.}
     \label{fig:license-share-ecosystems}
 \end{figure}

 \subsection{Feature analyses}

  \begin{figure}[tb]
      \centering
      \includegraphics[width=0.55\linewidth]{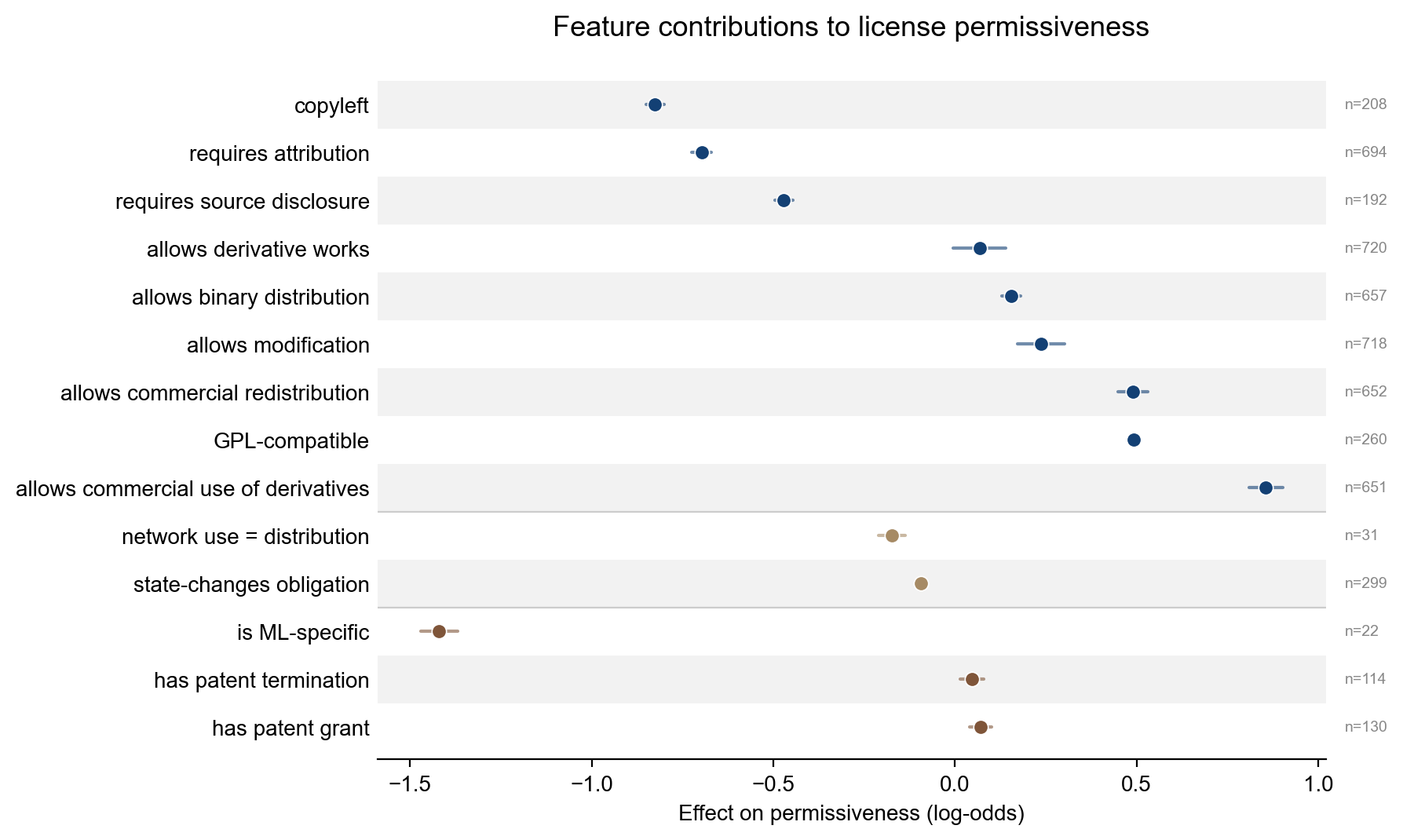}
      \caption{Features plotted based on logistic regression on permissiveness.}
      \label{fig:feature-regression-coverage}
  \end{figure}

When looking at the barbell plot in the main paper (Figure \ref{fig:feature_barbell}), we can study which individual features appear the most commonly and whether this signals permissiveness or restrictiveness. With a combinatorial view, we can do the same with the group of features that define a license. In this upset plot, we can visualize the features that differ and show the rate of divergence, where across all the pairs, we compare the less permissive license to the more permissive license, and see how many features differ and which ones. We can then count the rates of the different forms of divergence across all the instances.

  \begin{figure}
    \centering
    \includegraphics[width=0.46\linewidth]{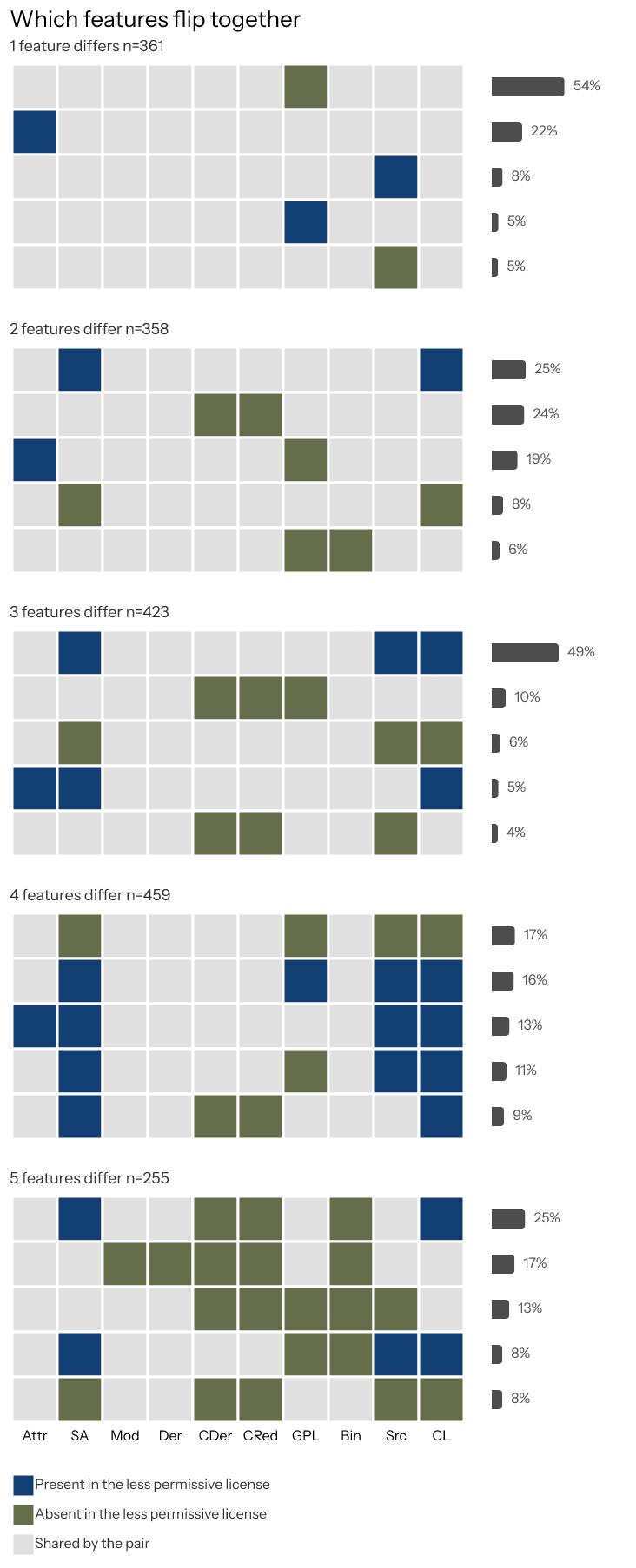}
    \caption{Upset plot depicting differences in ordered license pairs grouped
  by the number of features that differ \citep{2014_infovis_upset}. For
  every group, there is a bar plot showing the frequency of every divergence
  type. Gray cells mean that a feature is common between the pair. A blue
  cell means that given that this feature disagrees in a pair, it is because
  it was present in the less permissive license. An olive cell means that
  given this feature disagrees in a pair, it was not present in the less
  permissive license. This plot is a complement to \ref{tab:pairwise-comparisons}.}
    \label{fig:feature-diff}
\end{figure}

\end{document}